\documentclass[times]{elsarticle}
\usepackage{jcomp}
\usepackage{framed,multirow}

\usepackage{amssymb}
\usepackage{latexsym}
\usepackage{mathtools}
\usepackage{ucs}
\usepackage[utf8x]{inputenc}
\usepackage{url}

\definecolor{newcolor}{rgb}{.8,.349,.1}

\usepackage{graphicx}

\usepackage{amsmath}

\usepackage{enumerate}

\journal{Journal of Computational Physics}

\usepackage{graphicx}
\usepackage{amssymb}
\journal{Journal of Computational Physics}

\begin{document}

\begin{frontmatter}


\title{Gas Kinetic Schemes for Solving the Magnetohydrodynamic Equations with Pressure Anisotropy}



\author[1]{Hongyang Luo\corref{cor1}}
\author[2]{John G. Lyon}
\author[1,3]{Binzheng Zhang\corref{cor1}}

\address[1]{Department of Earth Sciences, the University of Hong Kong}
\address[2]{Department of Physics and Astronomy, Dartmouth College}
\address[3]{High Altitude Observatory, National Center for Atmospheric Research}

\cortext[cor1]{Corresponding Authors, H. Luo (hyluo@connect.hku.hk) and B. Zhang (binzh@hku.hk)}

\begin{abstract}

In many astrophysical plasmas, the Coulomb collision is insufficient to maintain an isotropic temperature, and the system is driven to the anisotropic regime. In this case, magnetohydrodynamic (MHD) models with anisotropic pressure are needed to describe such a plasma system. To solve the anisotropic MHD equation numerically, we develop a robust Gas-Kinetic flux scheme for non-linear MHD flows. Using anisotropic velocity distribution functions, the numerical flux functions are derived for updating the macroscopic plasma variables. The schemes is suitable for finite-volume solvers which utilize a conservative form of the mass, momentum and total energy equations, and can be easily applied to multi-fluid problems and extended to more generalized double polytropic plasma systems. Test results show that the numerical scheme is very robust and performs well for both linear wave and non-linear MHD problems. 

\end{abstract}


\end{frontmatter}

%\linenumbers

\section{Introduction}

The magnetohydrodynamics (MHD) theory plays an important role in studying various space and astrophysical plasma phenomena. While the ideal, isotropic MHD equations have been successfully applied to many plasma systems, e.g., the solar corona, the heliosphere and planetary magnetospheres, its validity is questionable since these collisionless space plasmas usually exhibit anisotropic temperature according to in-situ measurements \citep{paranicas1991,frank2004,matteini2007}. Thus anisotropic MHD theory is needed to describe such pressure anisotropy in collisionless plasma systems. Chew, Goldberger and Low (CGL) have derived the double-adiabatic theory for describing MHD flows with anisotropic pressures \citep{chew1956}. Assuming anisotropic velocity distribution functions, the moment integrals of the Vlasov equation gives the corresponding macroscopic equations for the perpendicular and parallel pressure with respect to the magnetic field. However, solving the CGL MHD equations numerically is very challenging since the equations are no longer fully conserved. Moreover, the magnitude of the pressure anisotropy also needs to be constrained since plasma instabilities are easily developed as the anisotropy approaches thresholds e.g., the firehose, mirror and ion-cyclotron instabilities. Such physical constraint are not fully described by the CGL MHD equations, and the treatment is likely problem-dependent. 

Wegmann \citep{wegmann1997} included anisotropic pressure in his one-fluid model, with a Godunov-type upwind difference scheme. Meng et al \citep{meng2012jcp,meng2012jgr} have developed numerical schemes for solving the anisotropic MHD equations based on applying the characteristic wave speeds of the CGL system in a Rusanov and/or HLL type flux function. To constrain the magnitude of the pressure anisotropy, a relaxation source term is introduced in the pressure equations based on the instability criteria. The scheme has been successfully used in complicated problems such as the terrestrial magnetosphere 
\citep{meng2012jgr} and the solar wind \citep{meng2015alfven}, showing promising improvements compared to the isotropic MHD models. Hirabayashi et al.\citep{hirabayashi2016} developed another scheme to solve for the anisotropic MHD equations, using a general pressure tensor with six distinct elements so no isotropic or gyrotropic assumpition is required. Similar to \citep{meng2012jcp,meng2012jgr}, numerical fluxes are calculated via the HLL method.  Test results have shown that the Hirabayashi et al schemes effectively handles both magnetized and
unmagnetized regions and properly reduces to both the isotropic and gyrotropic pressure approximations as asymptotes.

In general, solving the CGL MHD equations in a finite-volume framework requires the calculation of numerical flux at the cell interfaces to evolve the macroscopic fluid variables. Upwind schemes require calculations in the characteristic system, which can be quite complicated for anisotropic MHD equations. Central schemes are much simpler since no characteristic information is needed and approximate Riemann solvers can be used, e.g., the Rusanov solver\citep{rusanov1961} and the Harten-Lax-van Leer type solvers \citep{harten1983}, etc. On the other hand, Boltzmann schemes, also known as ``gas-kinetic schemes'', is another type of approximate Riemann solver that calculates the numerical fluxes across the interfaces by integrating the distribution functions over the velocity space\citep{croisille1995,xu1999}. This type of numerical techniques is examined to be very robust and reliable, especially on simplicity of of the kinetic flux functions, avoiding complicated wave decomposition procedure and entropy fix, and is adapted by the Lyon-Fedder-Mobarry (LFM) MHD code\citep{lyon2004} and the Grid Agnostic MHD for Extended Research Applications (GAMERA) code \citep{zhang2019}. Combined with a high-order reconstruction method, the gas-kinetic schemes used in the LFM MHD code is quite robust in various space plasma problems \citep{kallio1998venus,zhang2018jupiter}, and has been adapted to multi-fluid plasma problems \citep{brambles2011}. The GAMERA code is a reinvention of the LFM code with significant upgrades, and has successful applications in planetary modeling recently \citep{dang2022,zhang2021jupiter}.

In this paper, we extend the isotropic gas kinetic schemes by introducing temperature anisotropy in the microscopic distribution function of plasmas and derive the moment integrals to get macroscopic flux functions for advancing the MHD equations in a finite-volume framework. To ensure energy conservation when MHD shocks occur, we track the total energy and perpendicular pressure as the primary variables and derive the parallel pressure from the average scalar pressure. Combined with high-order reconstruction schemes, the new gas kinetic scheme is capable of solving MHD equations with anisotropic pressures. The scheme is accurate for linear wave problems and is robust for non-linear MHD flows such as strong shocks, and adapting to multi-fluid problems is straightforward. The paper is organized as follows: Section 2 describes governing equations of the model as well as a discussion of the instabilities. Section 3 presents the numerical method for the new gas-kinetic scheme. An example of extending the method to multi-dimensional applications is also shown in section 3. In section 4, numerical tests, including the Brio-Wu shock problem, one-dimensional magnetosonic wave, two-dimensional nonlinearly polarized circular Alfvén wave, Orszag–Tang Vortex as well as reconnection in the GEM challenge, are presented. We give a summary in section 5. An example one-dimensional Python code with the numerical technique described is also provided \citep{Luo2022}.

\section{The double-adiabatic(CGL) MHD equations}
\subsection{Governing Equations}

The conservative form of the double adiabatic equations can be written as follows:
\begin{equation}
    \frac{\partial \rho}{\partial t}=-\nabla \cdot(\rho \boldsymbol{u})
\end{equation}

\begin{equation}
    \frac{\partial \rho \boldsymbol{u}}{\partial t}=-\nabla \cdot(\rho \boldsymbol{u} \boldsymbol{u}+\overline{\boldsymbol{P}} )-\nabla \cdot\left(\overline{\boldsymbol{I}} \frac{B^{2}}{2}-\boldsymbol{B}\boldsymbol{B}\right) \label{momentumeq}
\end{equation}

\begin{equation}
    \frac{\partial \boldsymbol{B}}{\partial t}=-\nabla \times \boldsymbol{E},
\end{equation} 
where $\rho$ and $\boldsymbol{u}$ are plasma density and plasma bulk velocity, respectively.  $\boldsymbol{B}$ is the magnetic field, and $\boldsymbol{E}=-\boldsymbol{u} \times \boldsymbol{B}$ is the electric field based on the ideal Ohm's law. $\overline{\boldsymbol{P}}$ is the plasma thermal pressure tensor expressed as follows:

\begin{equation}
\overline{\boldsymbol{P}}=P_{\perp} \overline{\boldsymbol{I}}+\left(P_{\|}-P_{\perp}\right) \hat{\mathbf{b}} \hat{\mathbf{b}},
\label{ptensor}
\end{equation}
where  $\hat{\mathbf{b}}=\mathbf{B} /|\mathbf{B}|$ is the unit vector along the magnetic field, $P_{\|}$ and $P_{\perp}$ are the pressure components parallel and perpendicular to the magnetic field, respectively. Therefore the average scalar pressure can be then written as:
\begin{equation}
    P=\frac{2 P_{\perp}+P_{\|}}{3}
    \label{pconversion}
\end{equation}
which is one-third of the trace of the pressure tensor.
Without considering higher order moments (e.g., third moment heat fluxes), other than the ideal Faraday's Law, two adiabatic constants can be derived:

\begin{align}
\frac{D \frac{P_{\perp}}{\rho B}}{D t} &=0, \label{pperconv}\\
\frac{D \frac{P_{\|} B^{2}}{\rho^{3}}}{D t} &=0, \label{pparconv}
\end{align}
 where $D / D t=\partial / \partial t+\mathbf{u} \cdot \nabla$ is the Lagrangian derivative. Hau\citep{hau2002} showed that equations (\ref{pperconv}) and (\ref{pparconv}) can be put into conservative forms as follows:

\begin{align}
&\frac{\partial S_{\perp}}{\partial t}+\nabla \cdot\left(S_{\perp} \boldsymbol{u}\right)=0, \label{mueq} \\
&\frac{\partial S_{\|}}{\partial t}+\nabla \cdot\left(S_{\|} \boldsymbol{u}\right)=0 \label{sparaeq},
\end{align}
where $S_{\perp}=p_{\perp} B^{-1}$, $S_{\|}=p_{\|}(B / \rho)^{2}$ and $B=|\boldsymbol{B}|$ is the strength of the magnetic field. $S_{\perp}$ is the magnetic moment and will be notated as $\mu$ throughout the paper. More generalized double polytropic equations can be obtained by introducing appropriate polytropic exponents $\gamma_{\perp}$, $\gamma_{\|}$ with $S_{\perp}=p_{\perp}/ B^{\gamma{\perp}-1}$ and $S_{\|}=p_{\|}(B / \rho)^{\gamma_{\|}-1}$ \citep{hau1993double}. The double adiabatic equations can be interpreted as a limiting case with $\gamma_{\perp} = 2$ corresponding to degree of freedom $f=2$ and $\gamma_{\|} = 3$ corresponding to degree of freedom $f=1$. Note that the numerical method described in this paper can easily be extended to the generalized double polytropic cases since the double polytropic equations can also be casted into conservative form as Equations (\ref{mueq}) and (\ref{sparaeq}). 

To ensure energy conservation. We also solve for the plasma energy equation as used in previous MHD solvers \citep{lyon2004, zhang2019} for the average scalar pressure $P$ :

\begin{equation}
    \frac{\partial E_{P}}{\partial t}=-\nabla \cdot\left[\boldsymbol{u}\left(E_{P}+P\right)\right]-\boldsymbol{u} \cdot \nabla \cdot\left(\frac{B^{2}}{2} \overline{\boldsymbol{I}}-\boldsymbol{BB}\right) \label{eplasmaeq}
\end{equation}
where $E_{p}$ is the plasma energy, defined as follows:
\begin{equation}
 E_{p}=\frac{1}{2} \rho u^{2}+\frac{P}{\gamma-1}.
\end{equation}

The use of the plasma energy equation has significant advantages in a MHD flows with low plasma $\beta$. Although the total energy equation is a more proper choice for energy conservation, \citet{lyon2004} have shown that the use of plasma energy equation in numerical MHD does follow the Rankine–Hugoniot conditions within the truncation error, which is independent of whether or not the electric field is carried by dissipative processes through the shock.   Considering the energy conservation, jump condition near shock, and $\mu$ being a good constant of the motion, the plasma energy equation(\ref{eplasmaeq}) and the first invariant equation(\ref{mueq}) are used to determine the  the average pressure $P$ and perpendicular component $P_{\perp}$. The parallel pressure $P_{\|}$ is then calculated as $P_{\|} = 3P - 2P_{\perp}$ according to the equation(\ref{pconversion}). Nevertheless, solving for $P_{\|}$ using the second adiabat equation serves as a good check and the needed equations are also provided in the method derivation.

\subsection{Instabilities and relaxation}

In double-adiabatic MHD,  plasma instabilities occur due to strong pressure anisotropy. Physically, these instabilities tend to push the system to equilibrium and cause isotropizion of the plasma.  Without considering such isotropization processes, numerical solutions to the double-adiabatic MHD equation may lead to nonphysical results with pressure anisotropy exceeding the physical limits. To resolve the issue of non-physical pressure anisotropy, \citet{meng2012jcp,meng2012jgr} introduced a relaxation scheme using a operator splitting technique. The relaxation term is applied when any of the following instabilities criteria is reached:
\begin{align}
\frac{P_{\|}}{P_{\perp}}&>1+\frac{\mathbf{B}^{2}}{ P_{\perp}},\label{FHI}\\
\frac{P_{\perp}}{P_{\|}}&>1+\frac{\mathbf{B}^{2}}{2 P_{\perp}},\label{MIRI}\\
\frac{P_{\perp}}{P_{\|}}&>1+C_{1}\left(\frac{\mathbf{B}^{2}}{2  P_{\|}}\right)^{C_{2}},\label{ICI}
\end{align}
where (\ref{FHI}) describes the criterion for the firehose instability\citep{gary1998fhi}, (\ref{MIRI}) and (\ref{ICI}) correspond to the mirror instability and ion cyclotron instability\citep{gary1976proton,gary1992mirror}, respectively. $C_{1}$ and $C_{2}$ are constants depending on the field of interest and research approaches (e.g., \citet{anderson1994}, \citet{gary1994C1C2}). Here we use the set of values in \citet{meng2012jcp,meng2012jgr} with $C_{1} = 0.3$ and $C_{2} = 0.5$ for space plasma problems. Note that in the double-adiabatic MHD description of plasmas, only the firehose instability is resolved by the fluid assumption, while the mirror instability and ion cyclotron instability are of kinetic effects that cannot be captured by fluid model.

To impose limits on the pressure anisotropy from the numerical solutions, we use a similar relaxation method developed by \citet{meng2012jcp,meng2012jgr}.The basic idea of such relaxation is similar to that in \citep{denton2000}, which sets the distribution back to marginal stability. In our scheme, the relaxation process is applied on the perpendicular pressure $P_{\perp}$, while the parallel pressure $P_{\|}$ was used in \citet{meng2012jcp,meng2012jgr}. Thus the relaxation term in our calculation is expressed as:
\begin{equation}
\frac{\delta P_{\perp}}{\delta t}=\frac{\bar{P}_{\perp}-P_{\perp}}{\tau}    
\end{equation}
where $\bar{P}_{\perp}$ is the marginal stable value of the perpendicular pressure, obtained from Eqs.(\ref{pconversion}) as well as (\ref{FHI})-(\ref{ICI}). 
For example, if the firehose instability is present, the $\bar{P}_{\perp}$ is calculated from (\ref{FHI}) as followed:
\begin{equation}
    \bar{P}_{\perp} = P-\frac{{B}^{2}}{3}
\end{equation}

The marginal stable values for mirror instability and ion cyclotron instability are calculated through a similar process.  $\tau$ is the time rate at which $P_{\perp}$ approaches the marginal stable state, which can be either a constant value taken to be uniform in the simulation domain, or based on instabilities growth rate. Both approaches of determining $\tau$ should lead to much smaller value than the dynamical time of the system, and the results are compared in the application of geospace-type problem \citep{meng2012jgr}. With such technique, the pressure anisotropy is secure from reaching instabilities and breaking $\mu$ invariance. For now we adapt the first approach to set $\tau$. The relaxation term then can be applied in a point-implicit way, as a splitting operator at the end of each time step:

\begin{equation}
  P_{\perp}^{n+1}=P_{\perp}^{*}+\frac{\left(\bar{P}_{\perp}-P_{\perp}^{*}\right) \Delta t}{\Delta t+\tau}  
\end{equation}
where $\Delta t$ is the time step, $P_{\perp}^{*}$ and  $P_{\perp}^{n+1}$ are the perpendicular pressure value before and after the relaxation term is applied, respectively. If the pressure anisotropy exceeds thresholds of both mirror and ion cyclotron instabilities, the relaxation term with a larger value will be applied. We note that in global magnetospheric MHD models, besides a pressure relaxation term in unstable regions, a general global relaxation/isotropization term might be needed, as suggested in\citep{meng2012jgr}. Such global relaxation aims to represent other possible mechanisms restricting the plasma pressure anisotropy in the actual magnetosphere and will not be discussed here.

\section{Numerical Schemes}

\subsection{Fluid and Adiabatic Invariant Fluxes}

To compute the fluxes through cell interfaces for finite-volume solvers, we use a Boltzmann-type solver for the plasma part of the anisotropic MHD equations adapted from \citet{lyon2004}. Boltzmann solvers depend on integrating distribution functions with respect to the needed variables. The plasma distribution function can be a physical one, for example, describing the distribution of actual physical particles and their energy and momenta. It can also be more abstract, for example a function weighting the spread of Riemann invariants. In what follows, we use the common bi-Maxwellian distribution:
\begin{equation}
f\left(\mathbf{v}_{\perp}, \mathbf{v}_{\|}\right)=\left(\frac{1}{2 \pi}\right)^{3 / 2} \frac{1}{a^{2} b} \exp \left(\frac{-v_{\perp}^{2}}{2 a^{2}}+\frac{-v_{\|}^{2}}{2 b^{2}}\right),
\label{original unnormalized distribution}
\end{equation}
where $v_{\perp}$is two-dimensional in the two directions perpendicular to the magnetic field direction, which is arbitrary. $a = \left(P_{\perp}/\rho \right)^{1/2}$ and $b = \left(P_{\|}/\rho \right)^{1/2}$ are the perpendicular and parallel thermal speeds, respectively. In the following
 calculations, we use a unit normalization form for initial simplicity. Note that other forms of the distribution function may be used to derive the flux functions, following the same process as in the next sections.

\begin{figure}[htb!]
    \centering
    \includegraphics[width=0.65\textwidth]{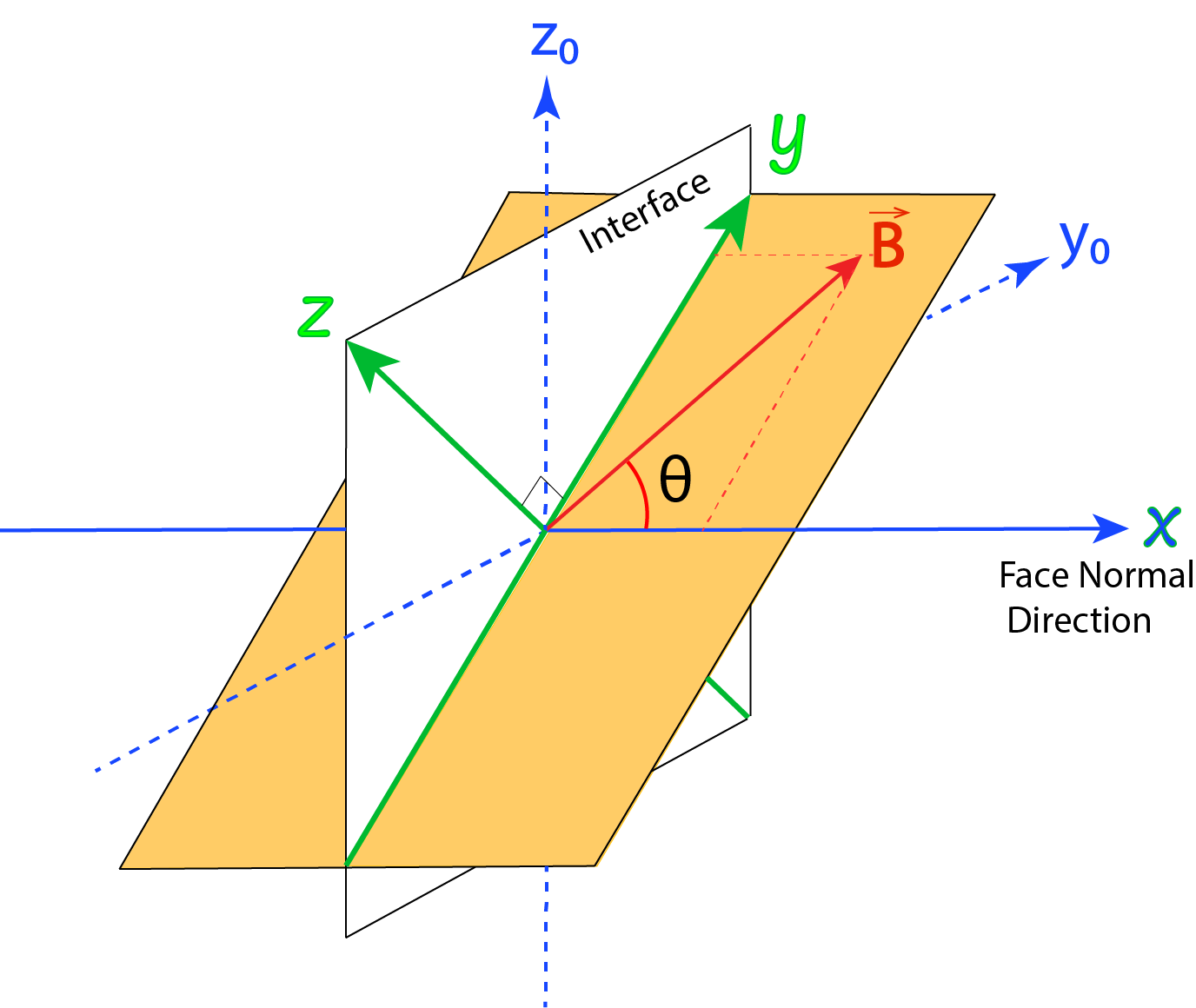}
    \caption{The interface coordinate system used in evaluating the numerical fluxes.}\label{coord}
\end{figure} 

In \citet{lyon2004}, the calculation of fluxes across a cell face is accomplished in a coordinate system fixed to the cell interface. Results are then transformed back to the global reference. By convention, here we use $\hat{\mathbf{x}}$ as the normal direction to the face. The other two direction vectors ($\hat{\mathbf{y}}_0$, $\hat{\mathbf{z}}_0$) are well-defined for the face and are consistent across the face, as shown in Fig \ref{coord}. For the calculation of anisotropic fluxes, it is convenient to perform a further transformation to a coordinate system that may be different on the two sides of the face, if the magnetic field differs across the interface. In the new coordinate system, $\hat{\mathbf{x}}$ remains the same, $\hat{\mathbf{z}}$ is defined by $\hat{\mathbf{x}} \times \mathbf{B}$, and $\hat{\mathbf{y}} = \hat{\mathbf{z}} \times \mathbf{x}$, forming an orthogonal right-handed Cartesian system. This amounts to a rotation about the original coordinate system so that $(x, y)$ plane contains the magnetic field, with $\mathbf{B}$ in the new system equal $B(\hat{\mathbf{x}} \cos \theta, \hat{\mathbf{y}} \sin \theta)$, as showm in Fig. \ref{coord}. Within the rotated interface coordinate system ($\hat{\mathbf{x}}$, $\hat{\mathbf{y}}$, $\hat{\mathbf{z}}$) the parallel and perpendicular velocities become
\begin{align}
\mathbf{v}_{\perp}&=-\hat{\mathbf{x}} v_{x} \sin \theta+\hat{\mathbf{y}} v_{y} \cos \theta+\hat{\mathbf{z}} v_{z}  &v_{\perp}^{2}&=v_{x}^{2} \sin ^{2}\theta-2 v_{x} v_{y} \cos \theta \sin \theta+v_{y}^{2} \cos ^{2}\theta+v_{z}^{2} \\
\mathbf{v}_{\|}&=\hat{\mathbf{x}} v_{x} \cos \theta+\hat{\mathbf{y}} v_{y} \sin \theta  &v_{\|}^{2}&=v_{x}^{2} \cos ^{2}\theta+2 v_{x} v_{y} \cos \theta \sin \theta+v_{y}^{2} \sin ^{2}\theta
\end{align}
It's useful to point out that the $\hat{\mathbf{b}} \hat{\mathbf{b}}$ dyadic is 
 \begin{equation}
  \hat{\mathbf{b}} \hat{\mathbf{b}}=\left(\begin{array}{ccc}\cos ^{2}\theta & \cos \theta \sin \theta & 0 \\ \sin \theta \cos \theta & \sin ^{2}\theta & 0 \\ 0 & 0 & 0\end{array}\right).
 \end{equation}
 
In terms of the $x, y, z$ velocities, the bi-Maxwellian distribution function (\ref{original unnormalized distribution}) becomes
\begin{equation}
f\left(v_{x}, v_{y}, v_{z}\right)=\left(\frac{1}{2 \pi}\right)^{3 / 2} \frac{1}{a^{2} b} \exp \left(\frac{-\left(a^{2}\left(v_{x}^{2}+v_{y} ^{2}\right)+\left(b^{2}-a^{2}\right)\left(v_{y} \cos \theta-v_{x} \sin \theta\right)^{2}\right)}{2 a^{2} b^{2}}+\frac{-v_{z}^{2}}{2 a^{2}}\right).
\end{equation}
 In general, $\mathbf{v}$ contains both the bulk velocity $\mathbf{u}$ and the thermal (peculiar) component $\mathbf{w}$. To simplify the calculation of the moment integrals, we transform the distribution function to a velocity system centered at the bulk velocity $\mathbf{u}$. The various moments of the Vlasov equation then become, for example:
\begin{equation}
M\left(v_{i}^{m} v_{j}^{n}\right)=\int_{-\infty}^{\infty}\left(u_{i}+w_{i}\right)^{m}\left(u_{j}+w_{j}\right)^{n} f(\mathbf{w}) d \mathbf{w}.
\end{equation}
In the $(x,y,z)$ coordinate system, $z$ integrals are separable, and the $y$ integrals can be evaluated in $[-\infty, \infty]$. Thus only the $x$ integrals need to be evaluated in a partial velocity domain. To evaluate the flux crossing a face, the rightward (positive) flux requires the integral of the distribution over $\left[-u_{x}, \infty\right]$ and the leftward over $\left[-\infty,-u_{x}\right]$. The separation into parallel and perpendicular velocities leaves cross terms, $w_{x} w_{y}$ in the exponential. These can be handled by completing the square in $w_{y} .$ The reduced distribution with y dependence removed is calculated as
 \begin{align}
 f_{0}^{(y)}\left(w_{x}\right) &=\frac{\exp \frac{-w_{x}^{2}}{2 \eta^{2}} }{ \sqrt{2 \pi}\eta} \\
\eta^{2} &=a^{2} \sin ^{2}\theta+b^{2} \cos ^{2}\theta  \quad=a^{2}+\left(b^{2}-a^{2}\right) \cos ^{2}\theta
\end{align}
The second form for $\eta$ shows the relationship to $P_{\|}-P_{\perp}$ that comes out later in the actual flux functions. We also need the first two moments of $w_{y}$ as functions of $w_{x}$.
\begin{align}
f_{1}^{(y)}\left(w_{x}\right)&=\frac{\exp \frac{-w_{x}^{2}}{2 \eta^{2}}\left(b^{2}-a^{2}\right) \cos \theta \sin \theta w_{x}}{\sqrt{2 \pi} \eta^{3}} \\
f_{2}^{(y)}\left(w_{x}\right)& =\frac{\exp \frac{-w_{x}^{2}}{2 \eta^{2}}\left(w_{x}^{2} \sin ^{2}\theta \cos ^{2}\theta\left(b^{2}-a^{2}\right)^{2}+a^{2} b^{2} \eta^{2}\right)}{\sqrt{2 \pi} \eta^{5}}
\end{align}

We define a number of integrals, denoted by $I_{m, n}^{L}$, where the superscript,L, denotes the rightward going integral $\int_{-u_{x}}^{\infty}$, i.e., flux from the left hand interface.  $m$ and $n$ refer to the powers of $w_{x}$ and $w_{y}$ in the integral moment. For example:
\begin{equation}
  I_{1,2}^{L}=\int_{-u_{x}}^{\infty} d w_{x} \int_{-\infty}^{\infty} d w_{y} w_{x}^{1} w_{y}^{2} f  
\end{equation}
and so on. Using the two-sided definition of the error function $erf(\cdot)$, i.e. $erf(0)=0$, $erf(-\infty)=-1$, and $erf(\infty)= 1$. The needed integrals are:
\begin{align}
I_{0,0}^{L} &=\frac{1-\operatorname{erf}\left(-\frac{u_{x}}{\sqrt{2} \eta}\right)}{2} \label{I28} \\ 
I_{1,0}^{L}&=\frac{\eta e^{\frac{-u_{x}^{2}}{2 \eta^{2}}}}{\sqrt{2 \pi}} \\
I_{2,0}^{L} &=\frac{\eta^{2}\left(1-\operatorname{erf}\left(-\frac{u_{x}}{\sqrt{2} \eta}\right)\right)-\sqrt{2 / \pi} \eta u_{x} e^{\frac{-u_{x}^{2}}{2 \eta^{2}}}}{2} \\
I_{3,0}^{L}&= \frac{ \eta\left(2 \eta^{2}+u_{x}^{2}\right) e^{\frac{-u_{x}^{2}}{2 \eta^{2}}}}{\sqrt{2\pi}} \\
I_{0,1}^{L} &=\frac{\left(b^{2}-a^{2}\right) \sin \theta \cos \theta e^{\frac{-u_{x}^{2}}{2 \eta^{2}}}}{\sqrt{2 \pi} \eta} \\
I_{1,1}^{L} &=\frac{1}{2}\left(b^{2}-a^{2}\right) \sin \theta \cos\theta\left(\left(1-\operatorname{erf}\left(-\frac{u_{x}}{\sqrt{2} \eta}\right)\right)-\frac{\sqrt{2} u_{x} e^{\frac{-u_{x}^{2}}{2 \eta^{2}}}}{\sqrt{\pi} \eta}\right)\\
I_{2,1}^{L}&=\frac{\sin \theta \cos \theta\left(b^{2}-a^{2}\right)\left(2 \eta^{2}+u_{x}^{2}\right) e^{\frac{-u_{x}^{2}}{2 \eta^{2}}}}{\sqrt{2 \pi} \eta} \\
I_{0,2}^{L}
&=\frac{1}{2 \sqrt{\pi} \eta^{3}}\left(-\sqrt{2}\left(b^{2}-a^{2}\right)^{2} \sin ^{2} \theta \cos ^{2} \theta u_{x} e^{\frac{-u_{x}^{2}}{2 \eta^{2}}} +
\notag\right.
\\
\phantom{=\;\;}
&\left.\quad
\left(1-\operatorname{erf}\left(\frac{-u_{x}^{2}}{\sqrt{2} \eta}\right)\right)\left(\left(b^{2}-a^{2}\right)^{2} \sin ^{2} \theta \cos ^{2} \theta+a^{2} b^{2}\right) \sqrt{\pi} \eta\right)\\
I_{1,2}^{L}&=\frac{1}{\sqrt{2 \pi} \eta^{3}} e^{\frac{-u_{x}^{2}}{2 \eta^{2}}}\left(a^{2} b^{2} \eta^{2}+\left(b^{2}-a^{2}\right)^{2} \sin ^{2} \theta \cos ^{2} \theta\left(2 \eta^{2}+u_{x}^{2}\right)\right) \label{I36}
\end{align} 

To show how these integrals align with the standard forms, they reduce to the following when the integral is runover the range, $[-\infty, \infty] .:$

\begin{flalign}
&\qquad \hspace{1.7cm}I_{0,0}=1 \\
&\qquad \hspace{1.7cm}I_{1,0}=0 \\
&\qquad \hspace{1.7cm}I_{2.0}=\eta^{2} \quad=a^{2}+\left(b^{2}-a^{2}\right) \cos ^{2}\theta\\
&\qquad \hspace{1.7cm}I_{3,0}=0 &
\end{flalign}

\begin{flalign}
&\qquad \hspace{1.7cm} I_{0,1}=0 \\
&\qquad \hspace{1.7cm} I_{1,1}=\left(b^{2}-a^{2}\right) \cos \theta \sin \theta \\
&\qquad \hspace{1.7cm} I_{2,1}=0 \\
&\qquad \hspace{1.7cm} I_{0,2}=a^{2}+\left(b^{2}-a^{2}\right) \sin ^{2}\theta \\
&\qquad \hspace{1.7cm} I_{1,2}=0&
\end{flalign} 

The leftward going integrals are $I^{R}=I-I^{L}$. Based on (\ref{I28})-(\ref{I36}), the rightward fluxes are calculated as:
\begin{align}
F^{L}(\rho)=& \int_{-\infty}^{\infty} d w_{y} \int_{-\infty}^{\infty} d w_{z} \int_{-u_{x}}^{\infty} d w_{x} v_{x} \rho^{L} f^{L}  \notag \\=& \int_{-\infty}^{\infty} d w_{y} \int_{-\infty}^{\infty} d w_{z} \int_{-u_{x}}^{\infty} d w_{x} \left( u_{x}+w_{x} \right) \rho^{L} f^{L}
=\rho^{L} \left(u_{x} I_{0,0}^{L}+I_{1,0}^{L} \right) \label{FrhoL}\\
F^{L}\left(p_{x}\right)=& \int_{-\infty}^{\infty} d w_{y} \int_{-\infty}^{\infty} d w_{z} \int_{-u_{x}}^{\infty} d w_{x} v_{x}^{2} \rho^{L} f^{L} \notag \\
=& \int_{-\infty}^{\infty} d w_{y} \int_{-\infty}^{\infty} d w_{z} \int_{-u_{x}}^{\infty} d w_{x} \left( u_{x}+w_{x} \right)^{2} \rho^{L} f^{L}=
 \rho^{L} \left( u_{x}^{2} I_{0,0}^{L}+2 u_{x} I_{1,0}^{L}+I_{2,0}^{L} \right) \\
F^{L}\left(p_{y}\right)=& \int_{-\infty}^{\infty} d w_{y} \int_{-\infty}^{\infty} d w_{z} \int_{-u_{x}}^{\infty} d w_{x} v_{x} v_{y} \rho^{L} f^{L} \notag \\
=& \int_{-\infty}^{\infty} d w_{y} \int_{-\infty}^{\infty} d w_{z} \int_{-u_{x}}^{\infty} d w_{x} \left( u_{x}+w_{x} \right)\left( u_{y}+w_{y} \right) \rho^{L} f^{L} \notag \\
=& \rho^{L} \left(u_{x} u_{y} I_{0,0}^{L}+u_{x} I_{0,1}^{L}+u_{y} I_{1,0}^{L}+I_{1,1}^{L}\right) \label{FpyL} \\
F^{L}\left(p_{z}\right)=&\int_{-\infty}^{\infty} d w_{y} \int_{-\infty}^{\infty} d w_{z} \int_{-u_{x}}^{\infty} d w_{x} v_{x} v_{z} \rho^{L} f^{L} \notag \\
=&\int_{-\infty}^{\infty} d w_{y} \int_{-\infty}^{\infty} d w_{z} \int_{-u_{x}}^{\infty} d w_{x} \left( u_{x}+w_{x} \right)\left( u_{z}+w_{z} \right) \rho^{L} f^{L} 
=\rho^{L} \left( u_{x} u_{z} I_{0,0}^{L} +u_{z} I_{1,0}^{L} \right) \label{FpzL} \\
F^{L}\left(E_{p l a s m a}\right)=&\frac{1}{2}\int_{-\infty}^{\infty} d w_{y} \int_{-\infty}^{\infty} d w_{z} \int_{-u_{x}}^{\infty} d w_{x} v_{x}\left( v_{x}^{2}+v_{y}^{2}+ v_{z}^{2}\right) \rho^{L} f^{L} \notag \\
=&\frac{1}{2}\int_{-\infty}^{\infty} d w_{y} \int_{-\infty}^{\infty} d w_{z} \int_{-u_{x}}^{\infty} d w_{x} v_{x} \left( \left( u_{x}+w_{x} \right)^{2}+\left( u_{y}+w_{y} \right)^{2}+ \left( u_{z}+w_{z} \right)^{2}\right) \rho^{L} f^{L} \notag \\
=& \frac{\rho^{L}}{2}\left(u_{x}\left(a^{2}+u_{x}^{2}+u_{y}^{2}+u_{z}^{2}\right) I_{0,0}^{L}+\left(a^{2}+3 u_{x}^{2}+u_{y}^{2}+u_{z}^{2}\right) I_{1,0}^{L}+3u_{x} I_{2,0}^{L}+I_{3,0}^{L}+
\notag\right.
\\
\phantom{=\;\;}
&\left.\quad 2u_{x} u_{y} I_{0,1}^{L}+u_{x} I_{0,2}^{L}+2u_{y} I_{1,1}^{L}+I_{1,2}^{L} \right) \label{FeplasmaL}
\\F^{L}(\mu)=&\frac{1}{2B^{L}} \int_{-\infty}^{\infty} d w_{y} \int_{-\infty}^{\infty} d w_{z} \int_{-u_{x}}^{\infty} d w_{x} v_{x} \left(\left(w_{x} \sin \theta+w_{y} \cos \theta\right)^2+w_{z}^2\right) \rho^{L} f^{L} \notag\\
=& \frac{\rho^{L}}{2B^{L}}\left(u_{x}\left(I_{0,0}^{L} a^{2}+I_{0,2}^{L} \cos ^{2}\theta-2 I_{1,1}^{L} \cos \theta \sin \theta+I_{2,0}^{L} \sin ^{2}\theta\right)+I_{1,0}^{L} a^2+I_{1,2}^{L} \cos ^{2}\theta \notag\right.\\
&\left.\quad-2 I_{2,1}^{L} \cos \theta \sin \theta+I_{3,0}^{L} \sin ^{2}\theta\right) \\
F^{L}\left(S_{\|} \right)=&\left(\frac{B^{L}}{\rho^{L}}\right)^{2} \int_{-\infty}^{\infty} d w_{y} \int_{-\infty}^{\infty} d w_{z} \int_{-u_{x}}^{\infty} d w_{x} v_{x} \left(w_{x} \cos \theta+w_{y} \sin \theta\right)^2 \rho^{L} f^{L} \notag \\ 
=&\left(\frac{B^{L}}{\rho^{L}}\right)^{2}\rho^{L}\left(u_{x}\left(I_{0,2}^{L} \sin ^{2}\theta+2 I_{1,1}^{L} \cos \theta \sin \theta+I_{2,0}^{L} \cos ^{2}\theta\right) \notag \right. \\
&\left.\hspace{1.2cm}+I_{1,2}^{L} \sin ^{2}\theta+2 I_{2,1}^{L} \cos \theta \sin \theta+I_{3,0}^{L} \cos ^{2}\theta\right)
\end{align}

The leftward going fluxes are the same with $I^{L}$ replaced with $I^{R}$. The fluxes $F$ at interface, is given by $F^{L}+F^{R}$, and if the state vectors are the same on both sides,  $F$ would be as follows:
\begin{align}
F(\rho) &=\rho u_{x} \label{Frhototal}\\
F\left(p_{x}\right) &=\rho \left( u_{x}^{2}+a^{2}+\left(b^{2}-a^{2}\right) \cos ^{2}\theta \right) \\
F\left(p_{y}\right) &=\rho \left(u_{x} u_{y}+\left(b^{2}-a^{2}\right) \sin \theta \cos \theta\right) \\
F\left(p_{z}\right) &=\rho u_{x} u_{z} \\
F\left(e_{p l a s m a}\right) &=\rho u_{x}\left(\left(u_{x}^{2}+u_{y}^{2}+u_{z}^{2}\right) / 2+\left(4 a^{2}+b^{2}\right) / 2+\left(b^{2}-a^{2}\right) \cos ^{2}\theta\right) \label{Feplasmatotal} \\
F(\mu) &=\frac{\rho}{B}\left( u_{x} a^2\right)\\
F\left(S_{\|}\right) &=\frac{B^{2}}{\rho}\left(u_{x} b^2 \right)
\end{align}

Set $ a=b=\sqrt{\frac{P}{\rho}}=\sqrt{\frac{P_{\perp}}{\rho}}=\sqrt{\frac{P_{\|}}{\rho}}$, (\ref{FrhoL})-(\ref{FeplasmaL}) and (\ref{Frhototal})-(\ref{Feplasmatotal}) recover the flux splitting schemes developed by \citet{xu1999} and used in \citet{zhang2019}.\par

\subsection{Magnetic Stresses}

To calculate the fluxes for magnetic stresses, we use a similar bi-Maxwellian distribution function with total pressure (gas+magnetic) for the values of $a$ and $b$. This choice of the distribution function is similar to the ones used in \citet{xu1999} and \citet{lyon2004} for computing the magnetic stresses, which has the mean speed within the distribution linked to the fast mode speed:
\begin{equation}
    f_{B}\left(\mathbf{v}_{\perp}, \mathbf{v}_{\|}\right)=\exp \left(\frac{-v_{\perp}^{2}}{2 a_{B}^{2}}+\frac{-v_{\|}^{2}}{2 b_{B}^{2}}\right),
\end{equation}
where $a_{B} = \sqrt{\frac{P_{ tot\perp}}{\rho}}$, $P_{ tot\perp} = P_{\perp}+\frac{1}{2}\left(B_{x}^{2}+B_{y}^{2}+B_{z}^{2}\right)$, and $b_{B} = \sqrt{\frac{P_{tot\|}}{\rho}}$, $P_{tot\|} = P_{\|}+\frac{1}{2}\left(B_{x}^{2}+B_{y}^{2}+B_{z}^{2}\right)$. Since the magnetic stress tensor does not explicitly contain the bulk velocity, only the zeroth moments of corresponding distribution that across the interface, i.e. $I_{B0,0}^{L}$ and $I_{B0,0}^{R}$ are needed, and  calculated as follows:
\begin{align}
I_{B0,0}^{L}  &=\int_{-u_{x}}^{\infty} d w_{x} \int_{-\infty}^{\infty} d w_{y} \int_{-\infty}^{\infty} d w_{z} f_{B}^{L} = \frac{1-\operatorname{erf}\left(-\frac{u_{x}}{\sqrt{2} \eta_{B}}\right)}{2} \\
I_{B0,0}^{R}  &=\int_{-\infty}^{-u_{x}} d w_{x} \int_{-\infty}^{\infty} d w_{y} \int_{-\infty}^{\infty} d w_{z} f_{B}^{R} = \frac{1+\operatorname{erf}\left(-\frac{u_{x}}{\sqrt{2} \eta_{B}}\right)}{2}, 
\end{align}
where $\eta_{B}= \sqrt{a_{B}^{2} \sin ^{2}\theta+b_{B}^{2} \cos ^{2}\theta} $. The magnetic stress tensor is calculated as follows:
\begin{equation}
\begin{aligned}
\overline{\boldsymbol{S}}_{\mathrm{mag}}
=&I_{B0,0}^{L}\left[\frac{1}{2}\left(B^{L}\right)^{2} \overline{\boldsymbol{I}}-\boldsymbol{B}^{L} \boldsymbol{B}^{L}\right] \\
&+I_{B0,0}^{R}\left[\frac{1}{2}\left(B^{R}\right)^{2} \overline{\boldsymbol{I}}-\boldsymbol{B}^{R} \boldsymbol{B}^{R}\right].
\end{aligned}
\end{equation}

\subsection{Coordinates Transforms to the Base System}

 So far everything is to have $x$ as the interface normal vector and the magnetic field is contained in the $(x,y)$ plane, which in general not aligned with the global reference $(x,y_0,z_0)$. To use these fluxes functions, it is convenient to define a rotated local coordinates ($x, y, z$) transformed from the original coordinate ($x_{0}, y_{0}, z_{0}$), after the left interface states are split to left and right states, then vector fluxes are solved and rotated back into the base system. 

\label{3D section}

\par One example of such transformation process, where $x$ is set to $x_{0}$-direction, is as follows:
\begin{equation}
\begin{aligned}
\boldsymbol{u}_{x,y,z} &=\overline{\boldsymbol{T}} \cdot \boldsymbol{u}_{x_{0}, y_{0}, z_{0}} \Rightarrow\left(\begin{array}{l}
u_{x} \\
u_{y} \\
u_{z}
\end{array}\right) \\
&=\left[\begin{array}{lll}
1 & \hspace{0.5cm} 0 & \hspace{0.3cm} 0\\
0 & \hspace{0.15cm}\cos \alpha & \sin \alpha \\
0  & -\sin \alpha &  \cos \alpha \\
\end{array}\right] \cdot\left(\begin{array}{l}
u_{x_{0}} \\
u_{y_{0}} \\
u_{z_{0}}
\end{array}\right),
\end{aligned}
\end{equation}
where $\cos \alpha = \frac{B_{y}+\epsilon}{\sqrt{B_{y}^{2}+B_{z}^{2}}+\epsilon} $, $\sin \alpha = \frac{B_{z}}{\sqrt{B_{y}^{2}+B_{z}^{2}}+\epsilon}$. The idea of introducing the infinitesimal term $\epsilon$ is optional, but it does account for including the special case that the direction of magnetic field is normal to the interface as well, i.e. B is aligned with $x$ and $\theta = 0$. In such case of $\theta = 0$, (\ref{FpyL}) and (\ref{FpzL}) are identical, therefore there is no need to distinguish $y$ from $z$. The inverse transformation matrix used to rotate the results back to global reference is simply the transpose of the tranformation matrix $\overline{\boldsymbol{T}}$ in that $\overline{\boldsymbol{T}}$ is a rotation matrix :
\begin{equation}
\begin{aligned}
\boldsymbol{F}_{\rho \boldsymbol{u}}^{x_{0}, y_{0}, z_{0}} &=\overline{\boldsymbol{T}}^{T} \cdot \boldsymbol{F}_{\rho \boldsymbol{u}}^{x,y,z} \Rightarrow\left(\begin{array}{l}
F_{x_{0}} \\
F_{y_{0}} \\
F_{z_{0}}
\end{array}\right) \\
&=\left[\begin{array}{lll}
1 & \hspace{0.3cm} 0 & \hspace{0.3cm} 0\\
0& \cos \alpha & -\sin \alpha  \\
0& \sin \alpha & \hspace{0.15cm}\cos \alpha 
\end{array}\right] \cdot\left(\begin{array}{l}
F_{x} \\
F_{y} \\
F_{z}
\end{array}\right).
\end{aligned}
\end{equation}
\par 
\par The transformation matrices for $y_{0}$-face-normal coordinate system and $z_{0}$-face-coordinate system go through the same process. 

\section{Test Results}

In this section, we show standard test simulation results to demonstrate the effectiveness of the anisotropic gas kinetic scheme for MHD problems, including both one-dimensional and two-dimensional MHD tests for both linear and nonlinear flow conditions. We use a similar finite-volume scheme as developed by \citet{zhang2019}, with high-order upwind reconstruction combined with the Partial Donor Cell (PDM) limiter, and constrained transport (Yee-Grid) to satisfy the divergence-free magnetic field $\nabla \cdot B = 0$. Since a second order Adams-Bashforth time stepping scheme is used in the test simulations, the relaxation terms serves as a splitting operator which is applied after the corrector step.

\subsection{1-D Linear Magnetosonic waves }

We first simulate the propagation of one-dimensional magnetosonic waves in the linear region with small velocity perturbation on a uniform background plasma and magnetic field. The simulated wave speeds are compared with the analytical solutions to demonstrate that the wave behavior follows the analytical dispersion relations.

The simulation domain is $x \in [-1,1]$ with $N_{x}=256$ grid cells. A hard-wall boundary condition is used in the simulation so that the linear wave exhibits standing-wave structures. The initial condition is set to $\rho = 1$, $P$ = 0.5, $V_{x} = 0.01 \sin 2\pi x$, $V_{y} = V_{z} = 0$, $B_{x} = B_{z} = 0$, and $B_{y} = 1$. The set of values of perpendicular and parallel pressures are then calculated according to specific anisotropy while keeping the average scalar pressure $P$ = 0.5. The initial magnitudes of the anisotropic pressure values used in the linear wave simulationss are listed in Table \ref{T0}.
\begin{table}[htb!]
\centering
\begin{tabular}{c|c|c|c|}
\cline{1-4}
\multicolumn{1}{|c|}{Pressure anisotropy $\frac{P_{\|}}{P_{\perp}}$}        & Parallel pressure $P_{\|}$  & Perpendicular pressure $P_{\perp}$  & Average scalar pressure $P$    \\ \hline
\multicolumn{1}{|c|}{0.25}  & 1/6 &  2/3  &  0.5         \\ \hline
\multicolumn{1}{|c|}{0.5}  & 0.3 & 0.6 & 0.5       \\ \hline
\multicolumn{1}{|c|}{1}  & 0.5 & 0.5 & 0.5   \\ \hline
\multicolumn{1}{|c|}{2}  & 0.75 & 0.375 & 0.5       \\ \hline
\multicolumn{1}{|c|}{3} & 0.9 & 0.3 & 0.5     \\ \hline
\multicolumn{1}{|c|}{4} & 1 & 0.25 & 0.5   \\ \hline

\end{tabular}
\caption{Values of pressure anisotropy $\frac{P_{\|}}{P_{\perp}}$, parallel pressure $P_{\|}$, perpendicular pressure $P_{\perp}$ and average scalar pressure $P$ in the presented test examples.}\label{T0}
\end{table}

\begin{figure}[bht!]
	\noindent\includegraphics[width=36pc]{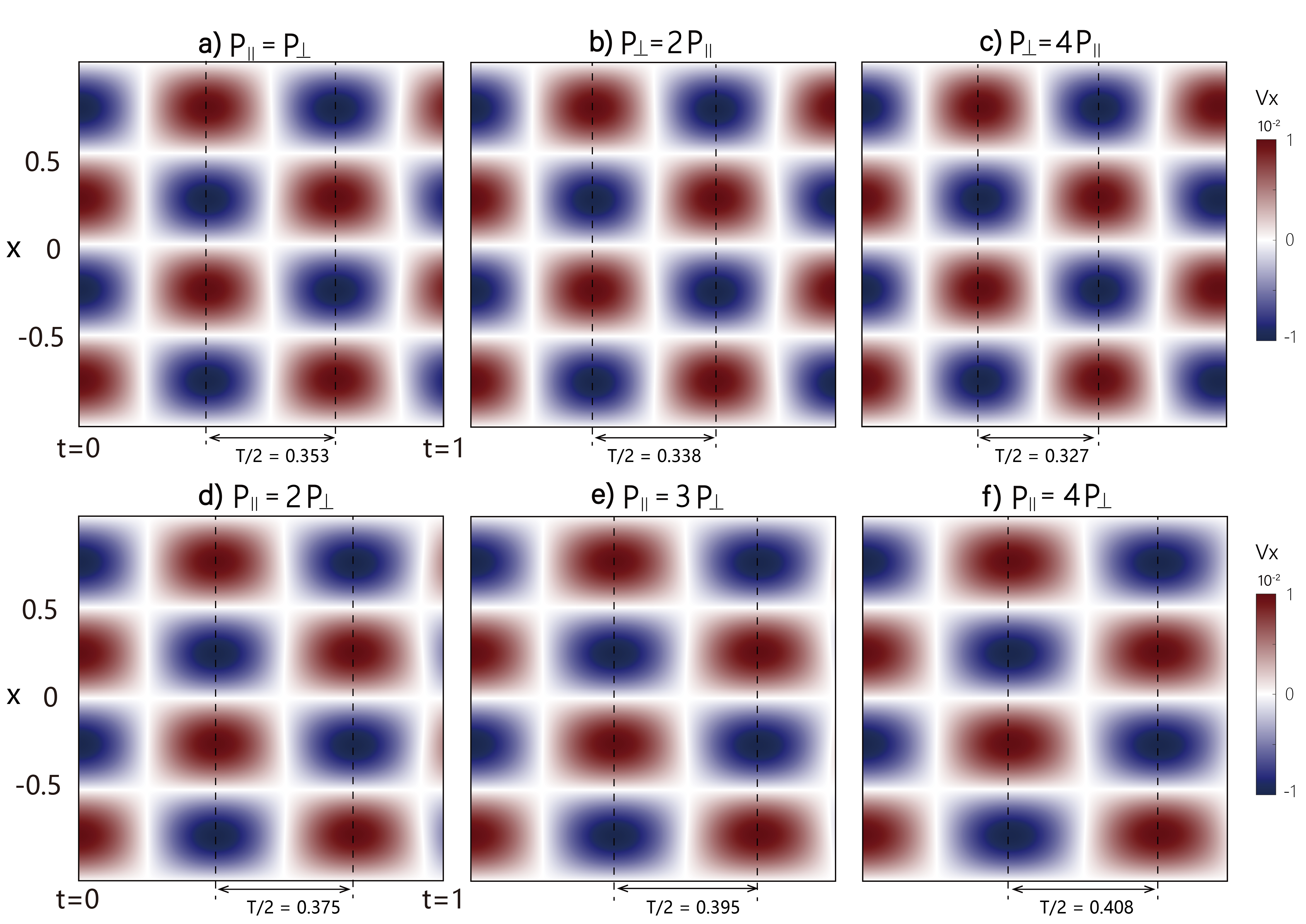}
	\centering
\caption{The $v_{x}$ as a function of time and position. The color shows the magnitude of $v_{x}$. The standing wave pattern is seen in the vertical (x) direction with the standing waves oscillations apparent in the horizontal (time). The oscillation period is shown by the arrowed indications beneath each plot.} \label{fastkeogram}
\end{figure}

\begin{figure}[htb!]
	\noindent\includegraphics[width=22pc]{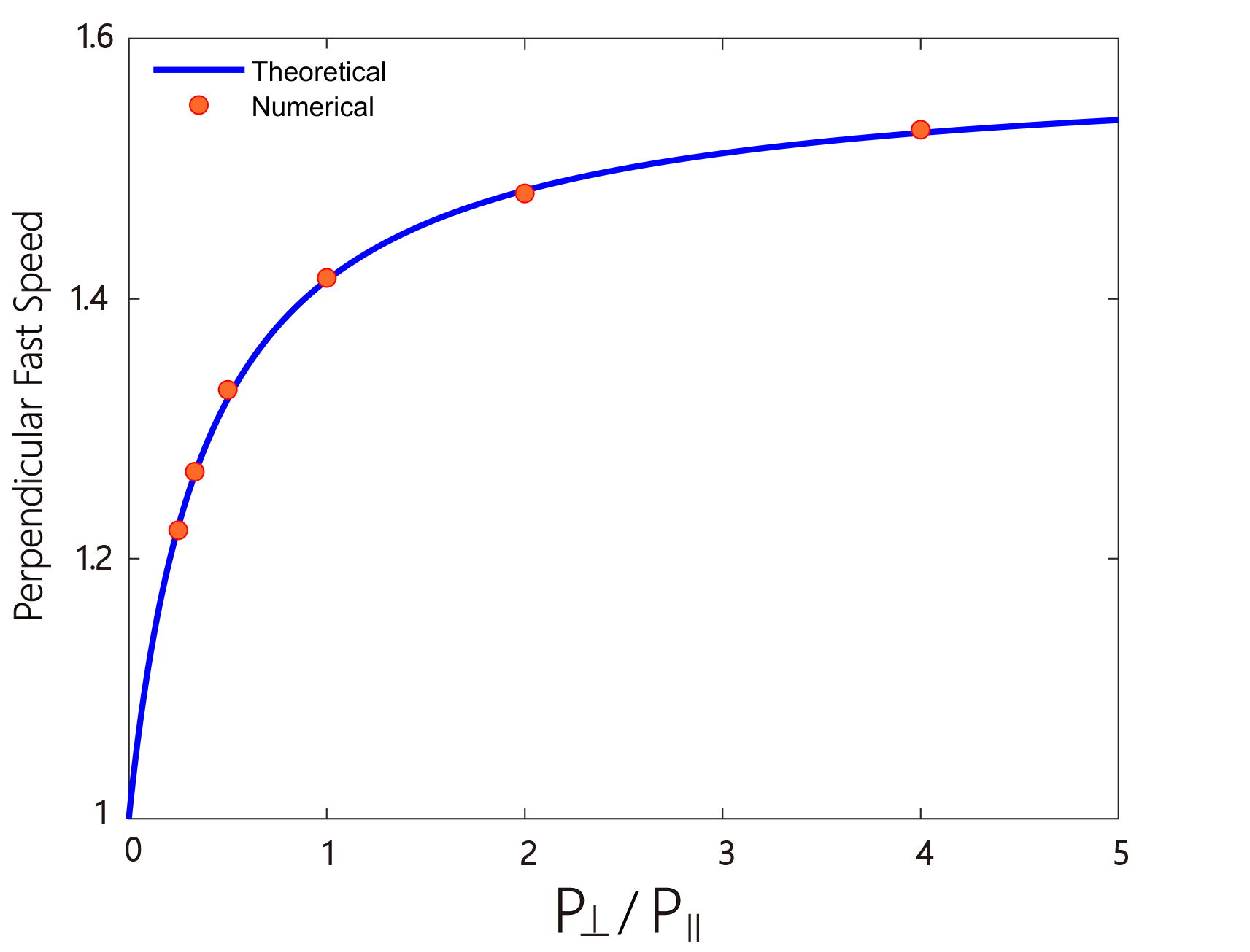}
	\centering
    \caption{Comparison of numerical and analytical perpendicular fast wave speed.}\label{fastcompare}
\end{figure}

The wave speed of the perpendicular fast mode is given by \citep{meng2012jcp}:
\begin{equation}
V_{F}=\sqrt{V_{A}^{2}+V_{S \perp}^{2}}=\sqrt{\frac{B^{2}}{\rho}+\frac{2p_{\perp}}{\rho}}
\end{equation}
To show the dynamic variation of the standing wave, we use a set of keograms showing the $v_{x}$ as a function of time and position under different pressure anisotropy as presented in Figure \ref{fastkeogram}. The phase speeds of the perpendicular magnetosonic modes are derived from the simulated periodicity, as shown in Figure \ref{fastkeogram}, which exhibit excellent agreement with the analytical wave speeds. The comparison of the numerical and theoretical values is shown in Figure \ref{fastcompare}.

\subsection{1-D MHD Shock Tube Problem}

To show the performace of the anisotropic gas kinetic scheme on handling non-linear problems with strong shocks, we use one-dimensional Brio-Wu shock tube problem as a standard test\citep{brio1988}. The 1-D MHD shock tube test is done in a domain of $x \in [-1,1]$ with $N_{x}=512$ cells. The initial conditions follow:
\begin{equation}
    \left(\rho, V_{x}, V_{y}, V_{z}, B_{x}, B_{y}, B_{z}, P, P_{\perp}, P_{\|}\right)= \begin{cases}(1.0,0.0,0.0,0.0,0.75,1.0,0.0,1.0,1.0,1.0) & (x < 0) \\ (0.125,0.0,0.0,0.0,0.75,-1.0,0.0,0.1,0.1,0.1) & (x\geq0)\end{cases}
\end{equation}

\begin{figure}[htb!]
    \noindent\includegraphics[width=40pc]{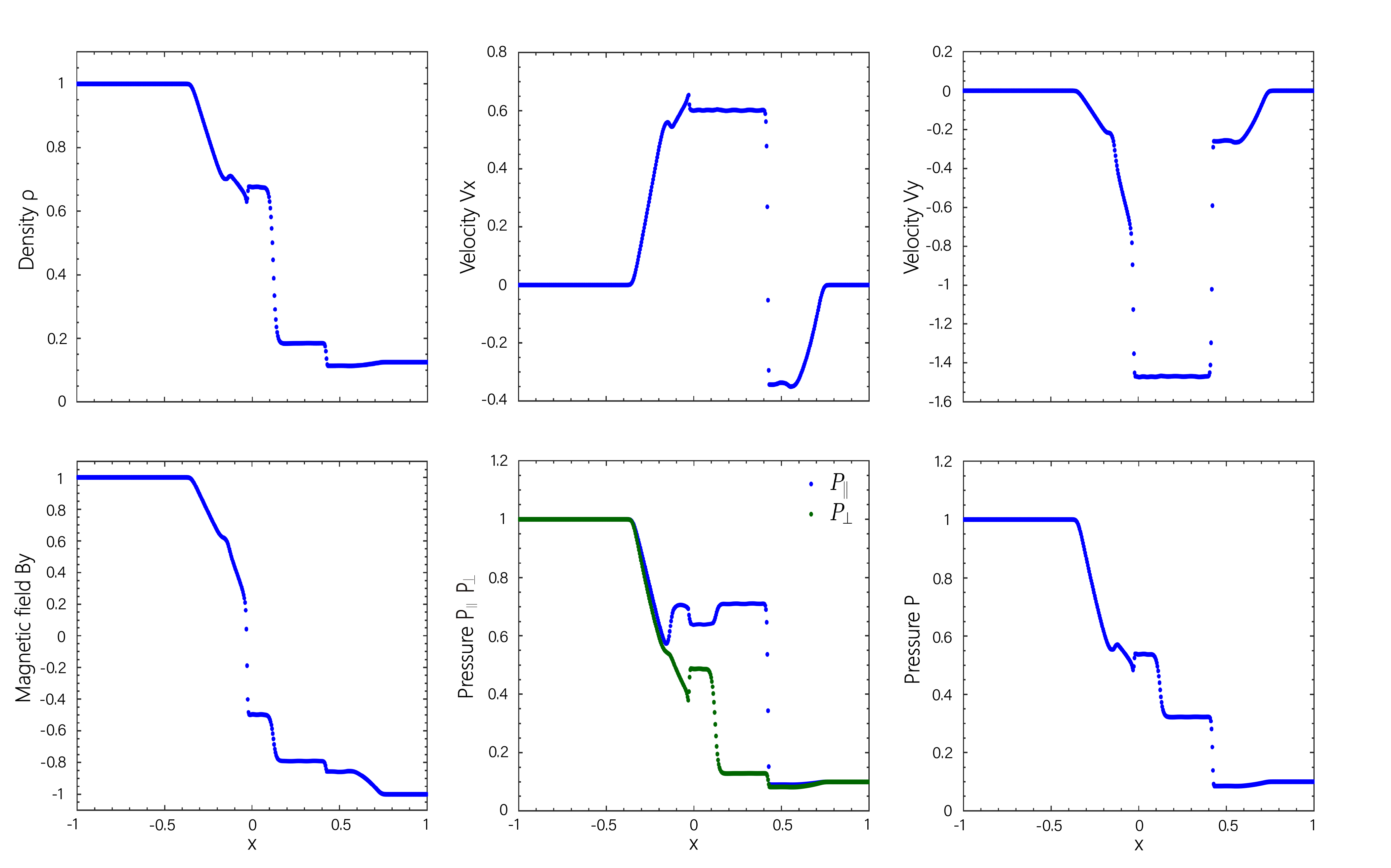}
	\centering
\caption{Brio–Wu shock tube problem under CGL MHD, data taken at t = 0.2.} \label{Briowu}
\end{figure}

The simulation results of the anisotropic MHD shock tube at $t=0.2$ are shown in Figure \ref{Briowu}. Similar test simulation results can be found in \citet{hirabayashi2016}. It is evident that our simulation results are very similar to Hirabayashi's nearly double adiabatic results. Both our results and those in \citet{hirabayashi2016} have shown: (1) the contact discontinuity region exhibits variations in $\rho, B_{y}, P_{\perp}$ and $P_{\|}$, which is different from the ideal MHD case, and (2) selective enhancement of the parallel pressure across the slow shock. The feature (1) can be explained by the momentum conservation law applied across a boundary without mass flux, and the feature (2) can be explained by the conservation of the two adiabatic invariants, i.e, Eqs.(\ref{pperconv})-(\ref{pparconv}). \citet{hirabayashi2016} provided detailed physical explanations for these noteworthy features compared with isotropic, ideal MHD solutions. We note that the jump condition of density and pressure is slightly different compared with \citet{hirabayashi2016}, probably because we solve the conservative form of the plasma energy equations, while the numerical schemes developed by \citet{hirabayashi2016} only used non-conservative form of the anisotropic pressure equations.

\subsection{2-D Nonlinearly Polarized Alfvén waves }

\begin{figure}[htb!]
    \noindent\includegraphics[width=22pc]{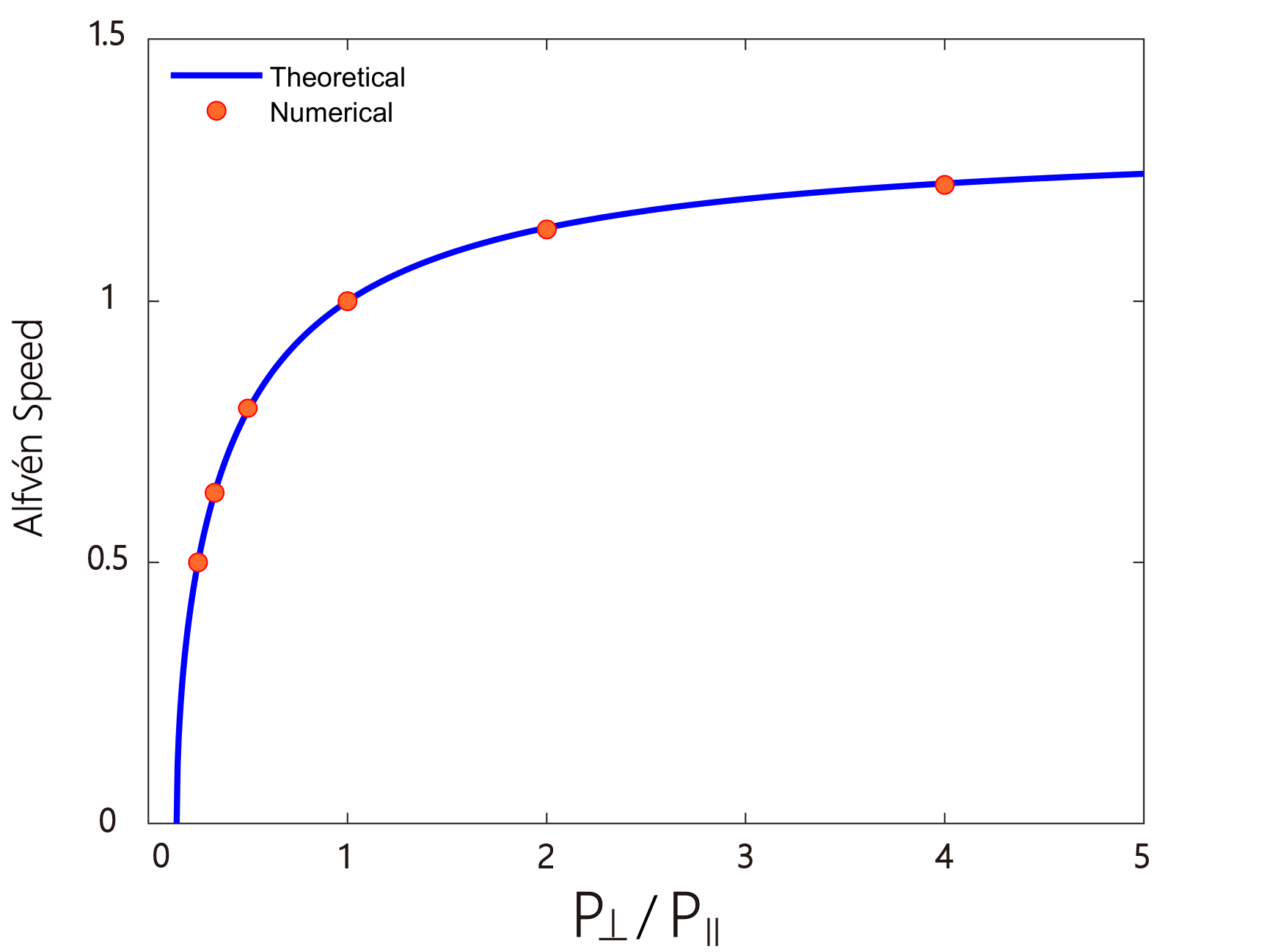}
    \centering
    \caption{Comparison of numerical and analytical Alfvén wave speed.}\label{Alfcompar}
\end{figure}

We use the nonlinearly polarized circular Alfvén wave test described in Tóth \citep{toth2000} to demonstrate the effectiveness of the new scheme in the nonlinear regime, as well as for multi-dimensional applications. The computational domain is set to $ 0 \leq x \leq \frac{1}{\sin{\alpha}}$ and $ 0 \leq y \leq \frac{1}{\cos{\alpha}}$,  where $\alpha$ = $\frac{\pi}{3}$ is the angle of Alfvén wave propagation with respect to the x-axis. The multi-dimensional nature of the test is guaranteed by having different numerical fluxes in the  x- and y-directions. Simulations are done with using a Cartesian grid with 128 $\times$ 128 cells, with periodic boundary conditions in both x- and y-directions. The initial conditions are $\rho=1$, $P = 0.5$, $u_{\perp}=\delta U \sin 2 \pi x_{\|}$, $B_{\perp} = \delta B \sin 2 \pi x_{\|}$, and $u_{z} = \delta U \cos 2 \pi x_{\|}$, $B_{z} = \delta B \cos 2 \pi x_{\|}$ with $\gamma=\frac{5}{3}$ and $x_{\|}=(x \cos \alpha+y \sin \alpha)$, where $u_{\perp}$ and $B_{\perp}$ are the components of the velocity and magnetic field perpendicular to the wave vector. The $B_{\|}$ and $B_{\perp}$ components are calculated as $B_{\perp}=B_{y} \cos \alpha-B_{x} \sin \alpha$, and $B_{\|}=$ $B_{x} \cos \alpha+B_{y} \sin \alpha$. The set of values of perpendicular and parallel pressures is the same as in the 1-D magnetosonic wave tests, shown in Table \ref{T0}. In order to make the non-linear Alfvén waves propagate in the direction $\alpha$,  the relation between $\delta U$ and $\delta B$ follows the Walen relation in anisotropic system, as suggested by \citet{hirabayashi2016}:
\begin{equation}
    \frac{\delta U}{V_{A}^{*}}=\frac{\delta B}{B_{\|0}},
\end{equation}
where $B_{\|0} = 1$ is the strength of the initial magnetic field parallel to the wave vector, $\delta B$ is set to 0.1, and $V_{A}^{*}$ is the modified Alfvén wave speed in anisotropic plasmas as follows:
\begin{equation}
    V_{A}^{*} = \sqrt{\frac{B^{2}+(P_{\perp}-P_{\|})}{\rho}},
    \label{VAequation}
\end{equation}
A set of keogram showing the $B_{\perp}$ as a function of time and position under different pressure anisotropy is presented in figure\ref{nonlin2D}. Compared with the isotropic case, the wave speed is faster when $p_\perp > p_\|$ and is slower when $p_\perp < p_\|$. A comparison of the numerical speed in the presented test cases and analytical propagation speed is shown in Figure \ref{Alfcompar}:

\begin{figure}[htb!]
	\noindent\includegraphics[width=36pc]{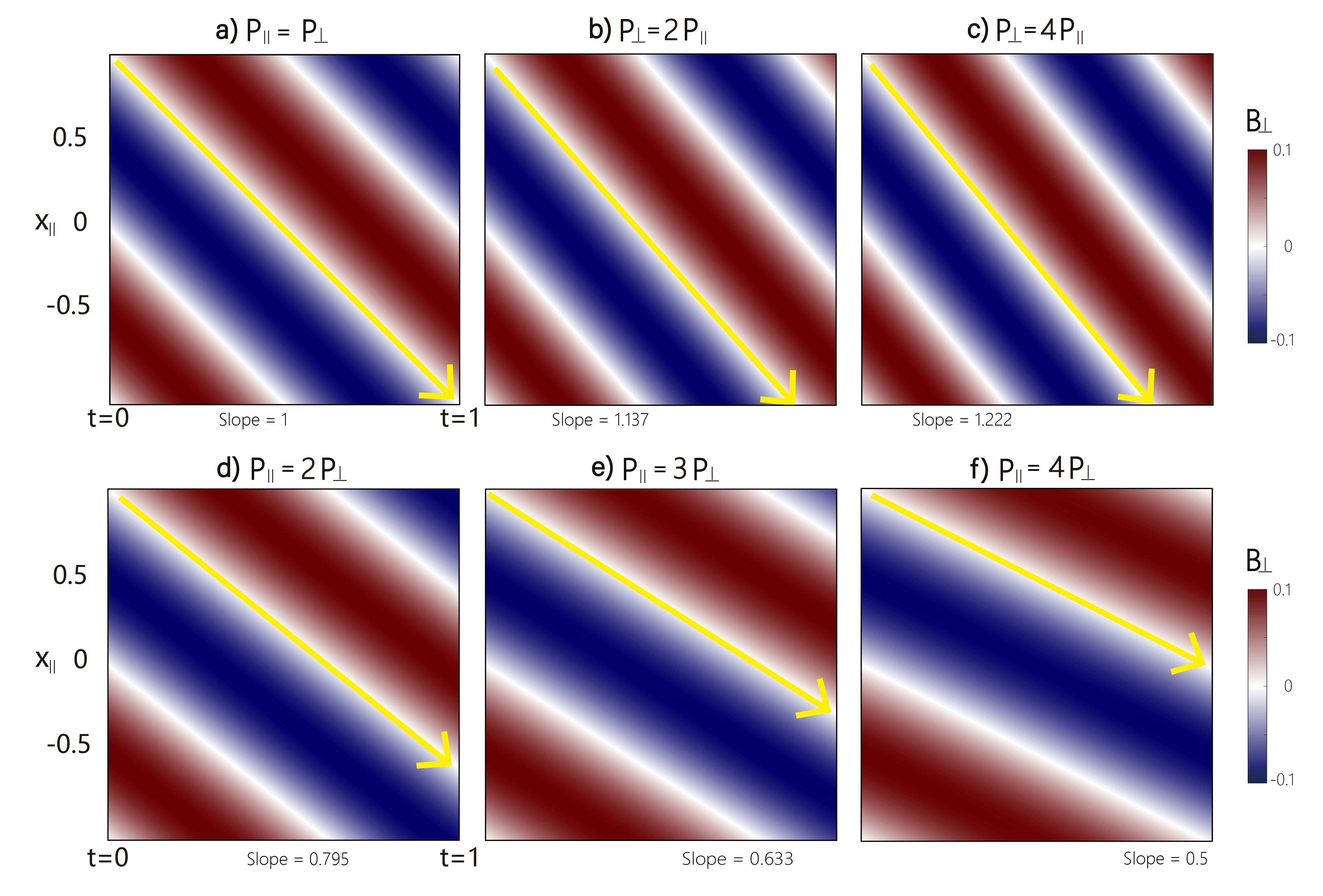}
	\centering
	\caption{The $B_{\perp}$ as a function of time and position. x-axis is time, y-axis is position ($x_{\|}$). The color shows the magnitude of $B_{\perp}$. The wave speed is shown by the slope of arrowed indications in each plot. Simulation time is 1.0. }\label{nonlin2D}
\end{figure}

\subsection{2-D Orszag–Tang Vortex}

To test the effectiveness of the anisotropic gas-kinetic scheme on tracking both discontinuities and smooth structures, we run the Orszag–Tang Vortex simulation \citep{orszag1979} using the double-adiabatic flux schemes. The test simulation is done within a square domain, with a grid of $x \in [0,1]$, $y \in [0,1]$, and $N_{x}=N_{y}=256$. The initial density and pressure are uniform within the simulation domain:  $\rho=\frac{25}{36} \pi$, $P=\frac{5}{12} \pi$, and $\gamma=\frac{5}{3}$. The initial velocities are set as periodic: $v_{x}=-\sin (2 \pi y)$ and $v_{y}=\sin (2 \pi x)$. The initial magnetic field are set as $B_{x}=-B_{0} \sin (2 \pi y)$ and $B_{y}=B_{0} \sin (4 \pi x)$ with $B_{0} = 1$. The boundary conditions are periodic in both x- and y-direction.

\begin{figure}[htb!]
	\noindent\includegraphics[width=32pc]{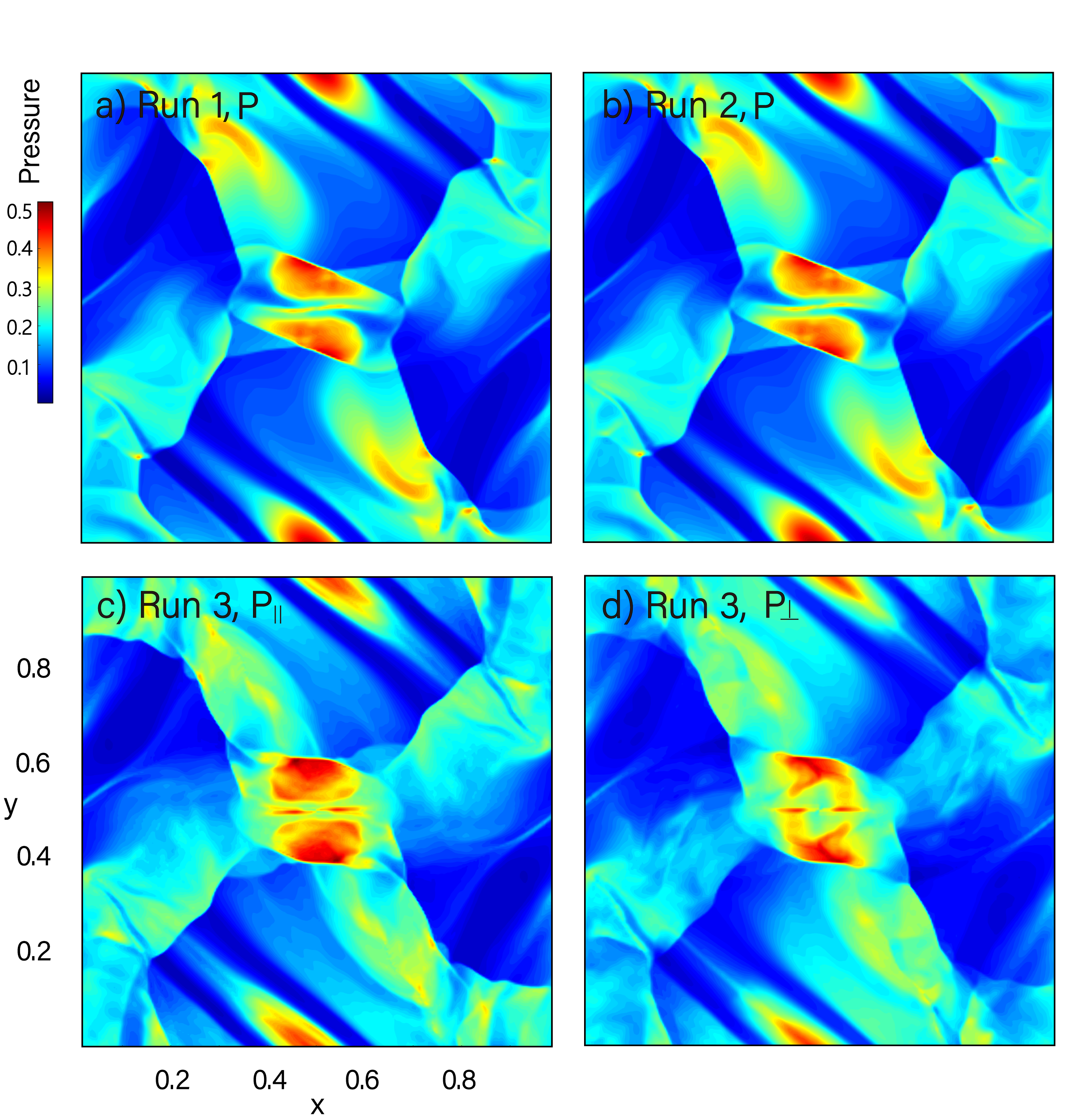}
	\centering
\caption{The spatial distribution of plasma pressure P at t = 0.48 in three Orszag–Tang simulations. Panel(a) shows the result in run 1 (ideal, isotropic model ). Panel(b) shows the result in run 2 (anisotropic model with enforced isotropization). Panel (c)(d) show $P_{\|}$ and $P_{\perp}$ in run 3 (anisotropic model), respectively. }\label{orzt plots}
\end{figure}

\begin{figure}[htb!]
	\noindent\includegraphics[width=26pc]{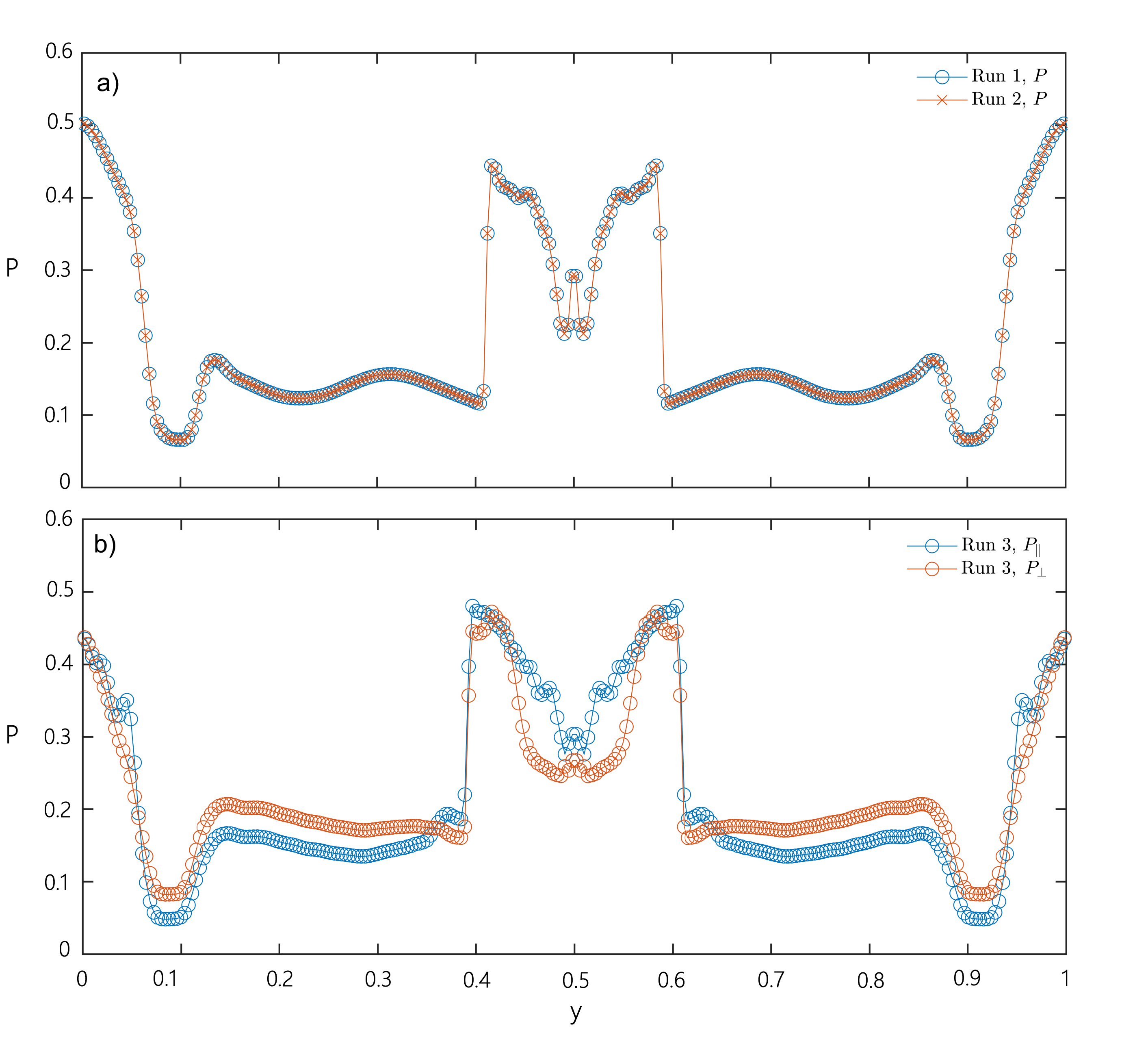}
	\centering

\caption{The line profiles of plasma pressure at x = 0.5 from three Orszag–Tang simulations at t = 0.48.}\label{orzt line profile}
\end{figure}

We perform three test simulations. Run 1 is from the isotropic, Maxwellian-based gas-kinetic scheme for ideal MHD, as developed by \citet{xu1999} and used in \citet{zhang2019}. Run 2 employs the anisotropic MHD scheme developed in this study, while isotropization is enforced, i.e., $P_{\|}=P_{\perp}=P$ at each time step. Run 3 uses the anisotropic MHD scheme, with relaxation time $\tau = 10^{-2}dt$ identical for three types of instabilities present in the computational domain, for simplicity. Figure \ref{orzt plots}(a) shows the spatial distributions of pressure at t = 0.48 in Run 1. Figure \ref{orzt plots}(b) shows the pressure from run 2 at the same simulation time. Figure \ref{orzt plots}(c) and (d) show the the spatial distributions of $P_{\|}$ and $P_{\perp}$ from Run 3 at $t=0.48$, respectively. The comparison between Figure \ref{orzt plots}(a) and (b) demonstrates that the anisotropic MHD scheme is reduced to the isotropic gas-kinetic scheme, when $P_{\|}=P_{\perp}=P$ is enforced. Note that up to simulation time t = 0.48, the overall structure of the anisotropic run 3 does not deviate from the ideal MHD result in an exaggerated/extreme way, since in the Orszag–Tang Vortex problem the plasma beta are large that  $\beta_{\|}$, $\beta_{\perp} > 1$, and hence the anisotropy is limited within a quite narrow range, by the instability condition. A more quantitative comparison of run 1 and run 2 is presented in Figure \ref{orzt line profile} using line profiles. Figure \ref{orzt line profile}(a) shows the comparisons of the plasma pressure
profiles (of run 1 and run 2) at t = 0.48, with x = 0.5, along the y direction. The simulated $P_{\|}$ and $P_{\perp}$ in Run 3 along the same x=0.5 cut line are presented in Figure\ref{orzt line profile}(b). The results show the effectiveness of the numerical scheme on handling the highly nonlinear MHD shock formation and interactions, as well as correctly reducing to isotropic scheme as a limiting case.

\subsection{The Geospace Environmental Modeling (GEM) Magnetic Reconnection Challenge}

We run the Geospace Environmental Modeling (GEM) Magnetic Reconnection Challenge \citep{birn2001summary,birn2001detail} to verify the scheme's capability of handling reconnection process. The initial conditions are a perturbed Harris sheet equilibrium. The unperturbed equilibrium is given by

\begin{align}
&B_{x} =B_0 \tanh (y / \lambda) \\
&n =n_0\left(1 / 5+\operatorname{sech}^2(y / \lambda)\right) \\
&P =\frac{B_0^2}{2 n_0} n(y), \\
\end{align}
And the perturbation is given as:
\begin{align}
\delta B &=-\hat{z} \times \nabla(\psi) \\
\psi(x, y) &=\psi_0 \cos \left(2 \pi x / L_x\right) \cos \left(\pi y / L_y\right),
\end{align}
where $\lambda =0.5,  B_0 =1,
n_0 =1$, $\psi_0 =B_0 / 10$, $L_x = 25.6$ and $L_y$ = 12.8. The boundary condition is periodic in the x-direction, and zero gradient is used in the y-direction. The 2-D computational domain is ranging from $x=+L_x/2$ to $x = -L_x/2$ and from $y = +L_y/2$ to $y = - L_y/2$, with $512 \times 256$ grid cells. Since our focus is to test the effectiveness of the anisotropic gas-kinetic schemes in an application like the GEM reconnection challenge, no resistive term is implemented in the test simulation, i.e., $\eta = 0$. In the simulation, no fast reconnection rate is observed since we did not include Hall physics. We also note that there is strong firehose-type anisotropy($P_{\|}-P_{\perp}$) in the outer layers of the magnetic islands but still inside 
the separatrix.  This distinguished feature is consistent with the observation in the anisotropic MHD result of \citep{birn2001detail}, and remains throughout the whole simulation, we give a snapshot of such feature at t=16 so the result can be compared with \citep{birn2001detail} Plate 4.

\begin{figure}[htb!]
	\noindent\includegraphics[width=32pc]{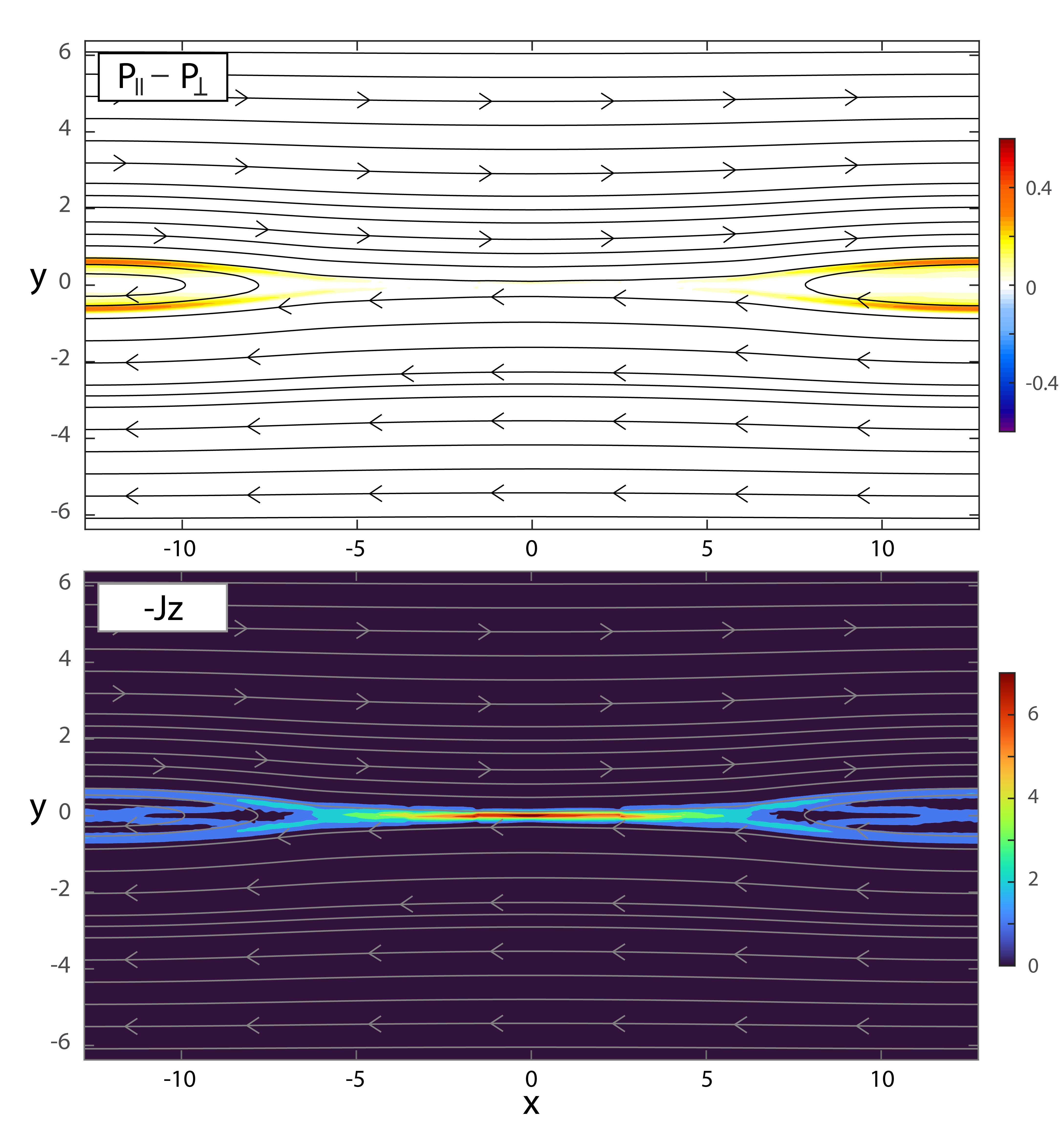}
	\centering

\caption{The pressure difference, current, as well as some magnetic field lines of the anisotropic run in the GEM reconnection at t = 16. Panel(a) shows the pressure difference, Panel(b) shows the current density Jz with a minus sign added to keep the direction of color bar consistent.}\label{GEM 16s}
\end{figure}

\section{Summary and Conclusion}
We proposed a new gas kinetic schemes for solving the double adiabatic MHD equations. The numerical method incorporates pressure anisotropy in the microscopic distribution function of plasmas. Moment integrals for macroscopic flux functions for conservative forms of anisotropic MHD equations (mass, momentum, energy as well as two adiabat), are derived. We implemented a source(relaxation) term to mimic micro-scale plasma interactions that relaxes the pressure to the marginally stable state, when the pressure anisotropy meets any instabilities criteria (fire-hose, mirror and ion cyclotron).

\par The numerical schemes have a comparable computational cost as the ideal MHD gas-kinetic flux splitting method \citep{xu1999},\citep{zhang2019} which has a few exp and erf function on each interface side. Since we use conservative form of pressure equations, the extension of the current numerical scheme to the generalized double polytropic equations is straightforward.

\par We perform a series of test cases to verify the numerical model. The results in both one-dimensional magnetosonic wave and two-dimensional nonlinearly polarized circular Alfvén wave propagation tests demonstrates the quality and accurateness of the current numeric scheme. 
The successful application in nonlinear test cases including Brio-Wu shock, Orszag–Tang Vortex and GEM reconnection simulations demonstrates the robustness of the method. We plan to apply the numerical model to geospace as well as planetary magnetospheres modeling. Extension to including Hall term and multi-fluid implementation will be done in the future.



%% References


\bibliographystyle{model1_num_names}
\bibliography{ref.bib}

\begin{thebibliography}{34}
\expandafter\ifx\csname natexlab\endcsname\relax\def\natexlab#1{#1}\fi
\providecommand{\bibinfo}[2]{#2}
\ifx\xfnm\relax \def\xfnm[#1]{\unskip,\space#1}\fi
%Type = Article
\bibitem[{Paranicas et~al.(1991)Paranicas, Mauk, and Krimigis}]{paranicas1991}
\bibinfo{author}{C.~Paranicas}, \bibinfo{author}{B.~Mauk},
  \bibinfo{author}{S.~Krimigis},
\newblock \bibinfo{title}{Pressure anisotropy and radial stress balance in the
  jovian neutral sheet},
\newblock \bibinfo{journal}{Journal of Geophysical Research: Space Physics}
  \bibinfo{volume}{96} (\bibinfo{year}{1991}) \bibinfo{pages}{21135--21140}.
%Type = Article
\bibitem[{Frank and Paterson(2004)}]{frank2004}
\bibinfo{author}{L.~Frank}, \bibinfo{author}{W.~Paterson},
\newblock \bibinfo{title}{Plasmas observed near local noon in jupiter's
  magnetosphere with the galileo spacecraft},
\newblock \bibinfo{journal}{Journal of Geophysical Research: Space Physics}
  \bibinfo{volume}{109} (\bibinfo{year}{2004}).
%Type = Article
\bibitem[{Matteini et~al.(2007)Matteini, Landi, Hellinger, Pantellini,
  Maksimovic, Velli, Goldstein, and Marsch}]{matteini2007}
\bibinfo{author}{L.~Matteini}, \bibinfo{author}{S.~Landi},
  \bibinfo{author}{P.~Hellinger}, \bibinfo{author}{F.~Pantellini},
  \bibinfo{author}{M.~Maksimovic}, \bibinfo{author}{M.~Velli},
  \bibinfo{author}{B.~E. Goldstein}, \bibinfo{author}{E.~Marsch},
\newblock \bibinfo{title}{Evolution of the solar wind proton temperature
  anisotropy from 0.3 to 2.5 au},
\newblock \bibinfo{journal}{Geophysical Research Letters} \bibinfo{volume}{34}
  (\bibinfo{year}{2007}).
%Type = Article
\bibitem[{Chew et~al.(1956)Chew, Goldberger, and Low}]{chew1956}
\bibinfo{author}{G.~Chew}, \bibinfo{author}{M.~Goldberger},
  \bibinfo{author}{F.~Low},
\newblock \bibinfo{title}{The boltzmann equation an d the one-fluid
  hydromagnetic equations in the absence of particle collisions},
\newblock \bibinfo{journal}{Proceedings of the Royal Society of London. Series
  A. Mathematical and Physical Sciences} \bibinfo{volume}{236}
  (\bibinfo{year}{1956}) \bibinfo{pages}{112--118}.
%Type = Article
\bibitem[{Wegmann(1997)}]{wegmann1997}
\bibinfo{author}{R.~Wegmann},
\newblock \bibinfo{title}{An upwind difference scheme for the double-adiabatic
  equations},
\newblock \bibinfo{journal}{Journal of Computational Physics}
  \bibinfo{volume}{131} (\bibinfo{year}{1997}) \bibinfo{pages}{199--215}.
%Type = Article
\bibitem[{Meng et~al.(2012{\natexlab{a}})Meng, T{\'o}th, Sokolov, and
  Gombosi}]{meng2012jcp}
\bibinfo{author}{X.~Meng}, \bibinfo{author}{G.~T{\'o}th},
  \bibinfo{author}{I.~V. Sokolov}, \bibinfo{author}{T.~I. Gombosi},
\newblock \bibinfo{title}{Classical and semirelativistic magnetohydrodynamics
  with anisotropic ion pressure},
\newblock \bibinfo{journal}{Journal of Computational Physics}
  \bibinfo{volume}{231} (\bibinfo{year}{2012}{\natexlab{a}})
  \bibinfo{pages}{3610--3622}.
%Type = Article
\bibitem[{Meng et~al.(2012{\natexlab{b}})Meng, T{\'o}th, Liemohn, Gombosi, and
  Runov}]{meng2012jgr}
\bibinfo{author}{X.~Meng}, \bibinfo{author}{G.~T{\'o}th},
  \bibinfo{author}{M.~Liemohn}, \bibinfo{author}{T.~Gombosi},
  \bibinfo{author}{A.~Runov},
\newblock \bibinfo{title}{Pressure anisotropy in global magnetospheric
  simulations: A magnetohydrodynamics model},
\newblock \bibinfo{journal}{Journal of Geophysical Research: Space Physics}
  \bibinfo{volume}{117} (\bibinfo{year}{2012}{\natexlab{b}}).
%Type = Article
\bibitem[{Meng et~al.(2015)Meng, Van~der Holst, T{\'o}th, and
  Gombosi}]{meng2015alfven}
\bibinfo{author}{X.~Meng}, \bibinfo{author}{B.~Van~der Holst},
  \bibinfo{author}{G.~T{\'o}th}, \bibinfo{author}{T.~Gombosi},
\newblock \bibinfo{title}{Alfv{\'e}n wave solar model (awsom): proton
  temperature anisotropy and solar wind acceleration},
\newblock \bibinfo{journal}{Monthly Notices of the Royal Astronomical Society}
  \bibinfo{volume}{454} (\bibinfo{year}{2015}) \bibinfo{pages}{3697--3709}.
%Type = Article
\bibitem[{Hirabayashi et~al.(2016)Hirabayashi, Hoshino, and
  Amano}]{hirabayashi2016}
\bibinfo{author}{K.~Hirabayashi}, \bibinfo{author}{M.~Hoshino},
  \bibinfo{author}{T.~Amano},
\newblock \bibinfo{title}{A new framework for magnetohydrodynamic simulations
  with anisotropic pressure},
\newblock \bibinfo{journal}{Journal of Computational Physics}
  \bibinfo{volume}{327} (\bibinfo{year}{2016}) \bibinfo{pages}{851--872}.
%Type = Article
\bibitem[{Rusanov(1961)}]{rusanov1961}
\bibinfo{author}{V.~V. Rusanov},
\newblock \bibinfo{title}{The calculation of the interaction of non-stationary
  shock waves with barriers},
\newblock \bibinfo{journal}{Zhurnal Vychislitel'noi Matematiki i
  Matematicheskoi Fiziki} \bibinfo{volume}{1} (\bibinfo{year}{1961})
  \bibinfo{pages}{267--279}.
%Type = Article
\bibitem[{Harten et~al.(1983)Harten, Lax, and Leer}]{harten1983}
\bibinfo{author}{A.~Harten}, \bibinfo{author}{P.~D. Lax},
  \bibinfo{author}{B.~v. Leer},
\newblock \bibinfo{title}{On upstream differencing and godunov-type schemes for
  hyperbolic conservation laws},
\newblock \bibinfo{journal}{SIAM review} \bibinfo{volume}{25}
  (\bibinfo{year}{1983}) \bibinfo{pages}{35--61}.
%Type = Article
\bibitem[{Croisille et~al.(1995)Croisille, Khanfir, and
  Chanteur}]{croisille1995}
\bibinfo{author}{J.-P. Croisille}, \bibinfo{author}{R.~Khanfir},
  \bibinfo{author}{G.~Chanteur},
\newblock \bibinfo{title}{Numerical simulation of the mhd equations by a
  kinetic-type method},
\newblock \bibinfo{journal}{Journal of scientific computing}
  \bibinfo{volume}{10} (\bibinfo{year}{1995}) \bibinfo{pages}{81--92}.
%Type = Article
\bibitem[{Xu(1999)}]{xu1999}
\bibinfo{author}{K.~Xu},
\newblock \bibinfo{title}{Gas-kinetic theory-based flux splitting method for
  ideal magnetohydrodynamics},
\newblock \bibinfo{journal}{Journal of Computational Physics}
  \bibinfo{volume}{153} (\bibinfo{year}{1999}) \bibinfo{pages}{334--352}.
%Type = Article
\bibitem[{Lyon et~al.(2004)Lyon, Fedder, and Mobarry}]{lyon2004}
\bibinfo{author}{J.~Lyon}, \bibinfo{author}{J.~Fedder},
  \bibinfo{author}{C.~Mobarry},
\newblock \bibinfo{title}{The lyon--fedder--mobarry (lfm) global mhd
  magnetospheric simulation code},
\newblock \bibinfo{journal}{Journal of Atmospheric and Solar-Terrestrial
  Physics} \bibinfo{volume}{66} (\bibinfo{year}{2004})
  \bibinfo{pages}{1333--1350}.
%Type = Article
\bibitem[{Zhang et~al.(2019)Zhang, Sorathia, Lyon, Merkin, Garretson, and
  Wiltberger}]{zhang2019}
\bibinfo{author}{B.~Zhang}, \bibinfo{author}{K.~A. Sorathia},
  \bibinfo{author}{J.~G. Lyon}, \bibinfo{author}{V.~G. Merkin},
  \bibinfo{author}{J.~S. Garretson}, \bibinfo{author}{M.~Wiltberger},
\newblock \bibinfo{title}{Gamera: A three-dimensional finite-volume mhd solver
  for non-orthogonal curvilinear geometries},
\newblock \bibinfo{journal}{The Astrophysical Journal Supplement Series}
  \bibinfo{volume}{244} (\bibinfo{year}{2019}) \bibinfo{pages}{20}.
%Type = Article
\bibitem[{Kallio et~al.(1998)Kallio, Luhmann, and Lyon}]{kallio1998venus}
\bibinfo{author}{E.~Kallio}, \bibinfo{author}{J.~Luhmann},
  \bibinfo{author}{J.~Lyon},
\newblock \bibinfo{title}{Magnetic field near venus: A comparison between
  pioneer venus orbiter magnetic field observations and an mhd simulation},
\newblock \bibinfo{journal}{Journal of Geophysical Research: Space Physics}
  \bibinfo{volume}{103} (\bibinfo{year}{1998}) \bibinfo{pages}{4723--4737}.
%Type = Article
\bibitem[{Zhang et~al.(2018)Zhang, Delamere, Ma, Burkholder, Wiltberger, Lyon,
  Merkin, and Sorathia}]{zhang2018jupiter}
\bibinfo{author}{B.~Zhang}, \bibinfo{author}{P.~Delamere},
  \bibinfo{author}{X.~Ma}, \bibinfo{author}{B.~Burkholder},
  \bibinfo{author}{M.~Wiltberger}, \bibinfo{author}{J.~Lyon},
  \bibinfo{author}{V.~Merkin}, \bibinfo{author}{K.~Sorathia},
\newblock \bibinfo{title}{Asymmetric kelvin-helmholtz instability at jupiter's
  magnetopause boundary: Implications for corotation-dominated systems},
\newblock \bibinfo{journal}{Geophysical Research Letters} \bibinfo{volume}{45}
  (\bibinfo{year}{2018}) \bibinfo{pages}{56--63}.
%Type = Article
\bibitem[{Brambles et~al.(2011)Brambles, Lotko, Zhang, Wiltberger, Lyon, and
  Strangeway}]{brambles2011}
\bibinfo{author}{O.~Brambles}, \bibinfo{author}{W.~Lotko},
  \bibinfo{author}{B.~Zhang}, \bibinfo{author}{M.~Wiltberger},
  \bibinfo{author}{J.~Lyon}, \bibinfo{author}{R.~Strangeway},
\newblock \bibinfo{title}{Magnetosphere sawtooth oscillations induced by
  ionospheric outflow},
\newblock \bibinfo{journal}{Science} \bibinfo{volume}{332}
  (\bibinfo{year}{2011}) \bibinfo{pages}{1183--1186}.
%Type = Article
\bibitem[{Dang et~al.(2022)Dang, Lei, Zhang, Zhang, Yao, Lyon, Ma, Xiao, Yan,
  Brambles et~al.}]{dang2022}
\bibinfo{author}{T.~Dang}, \bibinfo{author}{J.~Lei},
  \bibinfo{author}{B.~Zhang}, \bibinfo{author}{T.~Zhang},
  \bibinfo{author}{Z.~Yao}, \bibinfo{author}{J.~Lyon}, \bibinfo{author}{X.~Ma},
  \bibinfo{author}{S.~Xiao}, \bibinfo{author}{M.~Yan},
  \bibinfo{author}{O.~Brambles}, et~al.,
\newblock \bibinfo{title}{Oxygen ion escape at venus associated with
  three-dimensional kelvin-helmholtz instability},
\newblock \bibinfo{journal}{Geophysical Research Letters} \bibinfo{volume}{49}
  (\bibinfo{year}{2022}) \bibinfo{pages}{e2021GL096961}.
%Type = Article
\bibitem[{Zhang et~al.(2021)Zhang, Delamere, Yao, Bonfond, Lin, Sorathia,
  Brambles, Lotko, Garretson, Merkin et~al.}]{zhang2021jupiter}
\bibinfo{author}{B.~Zhang}, \bibinfo{author}{P.~A. Delamere},
  \bibinfo{author}{Z.~Yao}, \bibinfo{author}{B.~Bonfond},
  \bibinfo{author}{D.~Lin}, \bibinfo{author}{K.~A. Sorathia},
  \bibinfo{author}{O.~J. Brambles}, \bibinfo{author}{W.~Lotko},
  \bibinfo{author}{J.~S. Garretson}, \bibinfo{author}{V.~G. Merkin}, et~al.,
\newblock \bibinfo{title}{How jupiter’s unusual magnetospheric topology
  structures its aurora},
\newblock \bibinfo{journal}{Science Advances} \bibinfo{volume}{7}
  (\bibinfo{year}{2021}) \bibinfo{pages}{eabd1204}.
%Type = Article
\bibitem[{Luo et~al.(2022)Luo, Lyon, and Zhang}]{Luo2022}
\bibinfo{author}{H.~Luo}, \bibinfo{author}{J.~Lyon},
  \bibinfo{author}{B.~Zhang},
\newblock \bibinfo{title}{Gas kinetic schemes for solving the
  magnetohydrodynamic equations with pressure anisotropy},
\newblock \bibinfo{journal}{Zenodo. https://doi.org/10.5281/zenodo.7146168}
  (\bibinfo{year}{2022}).
%Type = Article
\bibitem[{Hau(2002)}]{hau2002}
\bibinfo{author}{L.-N. Hau},
\newblock \bibinfo{title}{A note on the energy laws in gyrotropic plasmas},
\newblock \bibinfo{journal}{Physics of Plasmas} \bibinfo{volume}{9}
  (\bibinfo{year}{2002}) \bibinfo{pages}{2455--2457}.
%Type = Article
\bibitem[{Hau et~al.(1993)Hau, Phan, Sonnerup, and Paschmann}]{hau1993double}
\bibinfo{author}{L.-N. Hau}, \bibinfo{author}{T.-D. Phan},
  \bibinfo{author}{B.~{\"O}. Sonnerup}, \bibinfo{author}{G.~Paschmann},
\newblock \bibinfo{title}{Double-polytropic closure in the magnetosheath},
\newblock \bibinfo{journal}{Geophysical research letters} \bibinfo{volume}{20}
  (\bibinfo{year}{1993}) \bibinfo{pages}{2255--2258}.
%Type = Article
\bibitem[{Gary et~al.(1998)Gary, Li, O'Rourke, and Winske}]{gary1998fhi}
\bibinfo{author}{S.~P. Gary}, \bibinfo{author}{H.~Li},
  \bibinfo{author}{S.~O'Rourke}, \bibinfo{author}{D.~Winske},
\newblock \bibinfo{title}{Proton resonant firehose instability: Temperature
  anisotropy and fluctuating field constraints},
\newblock \bibinfo{journal}{Journal of Geophysical Research: Space Physics}
  \bibinfo{volume}{103} (\bibinfo{year}{1998}) \bibinfo{pages}{14567--14574}.
%Type = Article
\bibitem[{Gary et~al.(1976)Gary, Montgomery, Feldman, and
  Forslund}]{gary1976proton}
\bibinfo{author}{S.~P. Gary}, \bibinfo{author}{M.~Montgomery},
  \bibinfo{author}{W.~Feldman}, \bibinfo{author}{D.~Forslund},
\newblock \bibinfo{title}{Proton temperature anisotropy instabilities in the
  solar wind},
\newblock \bibinfo{journal}{Journal of Geophysical Research}
  \bibinfo{volume}{81} (\bibinfo{year}{1976}) \bibinfo{pages}{1241--1246}.
%Type = Article
\bibitem[{Gary(1992)}]{gary1992mirror}
\bibinfo{author}{S.~P. Gary},
\newblock \bibinfo{title}{The mirror and ion cyclotron anisotropy
  instabilities},
\newblock \bibinfo{journal}{Journal of Geophysical Research: Space Physics}
  \bibinfo{volume}{97} (\bibinfo{year}{1992}) \bibinfo{pages}{8519--8529}.
%Type = Article
\bibitem[{Anderson et~al.(1994)Anderson, Fuselier, Gary, and
  Denton}]{anderson1994}
\bibinfo{author}{B.~J. Anderson}, \bibinfo{author}{S.~A. Fuselier},
  \bibinfo{author}{S.~P. Gary}, \bibinfo{author}{R.~E. Denton},
\newblock \bibinfo{title}{Magnetic spectral signatures in the earth's
  magnetosheath and plasma depletion layer},
\newblock \bibinfo{journal}{Journal of Geophysical Research: Space Physics}
  \bibinfo{volume}{99} (\bibinfo{year}{1994}) \bibinfo{pages}{5877--5891}.
%Type = Article
\bibitem[{Gary et~al.(1994)Gary, McKean, Winske, Anderson, Denton, and
  Fuselier}]{gary1994C1C2}
\bibinfo{author}{S.~P. Gary}, \bibinfo{author}{M.~E. McKean},
  \bibinfo{author}{D.~Winske}, \bibinfo{author}{B.~J. Anderson},
  \bibinfo{author}{R.~E. Denton}, \bibinfo{author}{S.~A. Fuselier},
\newblock \bibinfo{title}{The proton cyclotron instability and the
  anisotropy/$\beta$ inverse correlation},
\newblock \bibinfo{journal}{Journal of Geophysical Research: Space Physics}
  \bibinfo{volume}{99} (\bibinfo{year}{1994}) \bibinfo{pages}{5903--5914}.
%Type = Article
\bibitem[{Denton and Lyon(2000)}]{denton2000}
\bibinfo{author}{R.~E. Denton}, \bibinfo{author}{J.~G. Lyon},
\newblock \bibinfo{title}{Effect of pressure anisotropy on the structure of a
  two-dimensional magnetosheath},
\newblock \bibinfo{journal}{Journal of Geophysical Research: Space Physics}
  \bibinfo{volume}{105} (\bibinfo{year}{2000}) \bibinfo{pages}{7545--7556}.
%Type = Article
\bibitem[{Brio and Wu(1988)}]{brio1988}
\bibinfo{author}{M.~Brio}, \bibinfo{author}{C.~C. Wu},
\newblock \bibinfo{title}{An upwind differencing scheme for the equations of
  ideal magnetohydrodynamics},
\newblock \bibinfo{journal}{Journal of computational physics}
  \bibinfo{volume}{75} (\bibinfo{year}{1988}) \bibinfo{pages}{400--422}.
%Type = Article
\bibitem[{T{\'o}th(2000)}]{toth2000}
\bibinfo{author}{G.~T{\'o}th},
\newblock \bibinfo{title}{The∇{\textperiodcentered} b= 0 constraint in
  shock-capturing magnetohydrodynamics codes},
\newblock \bibinfo{journal}{Journal of Computational Physics}
  \bibinfo{volume}{161} (\bibinfo{year}{2000}) \bibinfo{pages}{605--652}.
%Type = Article
\bibitem[{Orszag and Tang(1979)}]{orszag1979}
\bibinfo{author}{S.~A. Orszag}, \bibinfo{author}{C.-M. Tang},
\newblock \bibinfo{title}{Small-scale structure of two-dimensional
  magnetohydrodynamic turbulence},
\newblock \bibinfo{journal}{Journal of Fluid Mechanics} \bibinfo{volume}{90}
  (\bibinfo{year}{1979}) \bibinfo{pages}{129--143}.
%Type = Article
\bibitem[{Birn et~al.(2001)Birn, Drake, Shay, Rogers, Denton, Hesse,
  Kuznetsova, Ma, Bhattacharjee, Otto et~al.}]{birn2001summary}
\bibinfo{author}{J.~Birn}, \bibinfo{author}{J.~Drake},
  \bibinfo{author}{M.~Shay}, \bibinfo{author}{B.~Rogers},
  \bibinfo{author}{R.~Denton}, \bibinfo{author}{M.~Hesse},
  \bibinfo{author}{M.~Kuznetsova}, \bibinfo{author}{Z.~Ma},
  \bibinfo{author}{A.~Bhattacharjee}, \bibinfo{author}{A.~Otto}, et~al.,
\newblock \bibinfo{title}{Geospace environmental modeling (gem) magnetic
  reconnection challenge},
\newblock \bibinfo{journal}{Journal of Geophysical Research: Space Physics}
  \bibinfo{volume}{106} (\bibinfo{year}{2001}) \bibinfo{pages}{3715--3719}.
%Type = Article
\bibitem[{Birn and Hesse(2001)}]{birn2001detail}
\bibinfo{author}{J.~Birn}, \bibinfo{author}{M.~Hesse},
\newblock \bibinfo{title}{Geospace environment modeling (gem) magnetic
  reconnection challenge: Resistive tearing, anisotropic pressure and hall
  effects},
\newblock \bibinfo{journal}{Journal of Geophysical Research: Space Physics}
  \bibinfo{volume}{106} (\bibinfo{year}{2001}) \bibinfo{pages}{3737--3750}.

\end{thebibliography}






% \end{thebibliography}

\end{document}